\documentclass[fleqn,usenatbib]{mnras}

\usepackage{newtxtext,newtxmath}

\usepackage[T1]{fontenc}
\usepackage{subcaption}
\usepackage{booktabs}
\usepackage{multirow}

\DeclareRobustCommand{\VAN}[3]{#2}
\let\VANthebibliography\thebibliography
\def\thebibliography{\DeclareRobustCommand{\VAN}[3]{##3}\VANthebibliography}

\usepackage{graphicx}	
\usepackage{amsmath}	

\usepackage{xcolor,bm}
\usepackage{xspace}
\usepackage{hyperref}
\usepackage{footnote}
\usepackage{footmisc}
\usepackage{orcidlink}

\newcommand{\lya}{Ly$\alpha$\xspace}
\newcommand{\lyb}{Ly$\beta$\xspace}

\title[]{Modelling the impact of quasar redshift errors on the full-shape analysis of correlations in the Lyman-$\alpha$ forest.}

\author[C. Gordon et al.]{Calum Gordon \orcidlink{0000-0003-2561-5733},$^{1}$\thanks{E-mail: cgordon@ifae.es}
Andrei Cuceu \orcidlink{0000-0002-2169-0595},$^{2,3}$\thanks{NASA Einstein Fellow}
Andreu Font-Ribera \orcidlink{0000-0002-3033-7312},$^{1}$
Hiram K. Herrera-Alcantar \orcidlink{0000-0002-9136-9609},$^{3,4}$
\newauthor
Jessica Nicole Aguilar,$^{2}$
Steven Ahlen \orcidlink{0000-0001-6098-7247},$^{5}$
Davide Bianchi \orcidlink{0000-0001-9712-0006},$^{6,7}$
David Brooks,$^{8}$
Todd Claybaugh,$^{2}$
\newauthor
Shaun Cole \orcidlink{0000-0002-5954-7903},$^{9}$
Axel de la Macorra \orcidlink{0000-0002-1769-1640},$^{10}$
Biprateep Dey \orcidlink{0000-0002-5665-7912},$^{11,12}$
Peter Doel,$^{8}$
Jaime E. Forero-Romero \orcidlink{0000-0002-2890-3725},$^{13,14}$
\newauthor
Enrique Gaztañaga,$^{15,16,17}$
Satya Gontcho A Gontcho \orcidlink{0000-0003-3142-233X},$^{2}$
Gaston Gutierrez,$^{18}$
Julien Guy \orcidlink{0000-0001-9822-6793},$^{2}$
\newauthor
Klaus Honscheid \orcidlink{0000-0002-6550-2023},$^{19,20,21}$
Mustapha Ishak \orcidlink{0000-0002-6024-466X},$^{22}$
Robert Kehoe,$^{23}$
David Kirkby \orcidlink{0000-0002-8828-5463},$^{24}$
Theodore Kisner \orcidlink{0000-0003-3510-7134},$^{2}$
\newauthor
Anthony Kremin \orcidlink{0000-0001-6356-7424},$^{2}$
Martin Landriau \orcidlink{0000-0003-1838-8528},$^{2}$
Laurent Le Guillou \orcidlink{0000-0001-7178-8868},$^{25}$
Michael Levi \orcidlink{0000-0003-1887-1018},$^{2}$
Marc Manera \orcidlink{0000-0003-4962-8934},$^{26,1}$
\newauthor
Paul Martini \orcidlink{0000-0002-4279-4182},$^{19,27,21}$
Ramon Miquel,$^{28,1}$
John Moustakas \orcidlink{0000-0002-2733-4559},$^{29}$
Seshadri Nadathur \orcidlink{0000-0001-9070-3102},$^{16}$
Gustavo Niz \orcidlink{0000-0002-1544-8946},$^{30,31}$
\newauthor
Nathalie Palanque-Delabrouille \orcidlink{0000-0003-3188-784X},$^{4,2}$
Will Percival \orcidlink{0000-0002-0644-5727},$^{32,33,34}$
Francisco Prada \orcidlink{0000-0001-7145-8674},$^{35}$
Ignasi Pérez-Ràfols \orcidlink{0000-0001-6979-0125},$^{36}$
\newauthor
Graziano Rossi,$^{37}$
Eusebio Sanchez \orcidlink{0000-0002-9646-8198},$^{38}$
David Schlegel,$^{2}$
Michael Schubnell,$^{39,40}$
Hee-Jong Seo \orcidlink{0000-0002-6588-3508},$^{41}$
\newauthor
Joseph Harry Silber \orcidlink{0000-0002-3461-0320},$^{2}$
David Sprayberry,$^{42}$
Gregory Tarlé \orcidlink{0000-0003-1704-0781},$^{40}$
Benjamin Alan Weaver,$^{42}$
Rongpu Zhou \orcidlink{0000-0001-5381-4372},$^{2}$
\newauthor
Hu Zou \orcidlink{0000-0002-6684-3997}$^{43}$ \\ \\ \textit{Affiliations are listed at the end of the paper.}}

\date{Accepted XXX. Received YYY; in original form ZZZ}

\pubyear{2025}

\begin{document}
\label{firstpage}
\pagerange{\pageref{firstpage}--\pageref{lastpage}}
\maketitle

\begin{abstract}
In preparation for the first cosmological measurements from the full-shape of the Lyman-$\alpha$ (\lya) forest from DESI, we must carefully model all relevant systematics that might bias our analysis. It was shown in \cite{Youles} that random quasar redshift errors produce a smoothing effect on the mean quasar continuum in the \lya forest region. This in turn gives rise to spurious features in the \lya auto-correlation, and its cross-correlation with quasars. Using synthetic data sets based on the DESI survey, we confirm that the impact on BAO measurements is small, but that a bias is introduced to parameters which depend on the full-shape of our correlations. We combine a model of this contamination in the cross-correlation \citep{Youles} with a new model we introduce here for the auto-correlation. These are parametrised by 3 parameters, which when included in a joint fit to both correlation functions, successfully eliminate any impact of redshift errors on our full-shape constraints. We also present a strategy for removing this contamination from real data, by removing $\sim$0.3\% of correlating pairs.

\end{abstract}

\begin{keywords}
(cosmology:) large-scale structure of Universe -- (cosmology:) distance scale -- (cosmology:) dark energy
\end{keywords}

\section{Introduction}\label{sec:intro}

One of the major objectives in modern cosmology is to understand the nature of dark energy. The discovery of the accelerating expansion of the universe from measurements of distant supernovae \citep{Riess98,Perlmutter99} introduced the need for some form of dark energy, in the simplest case a cosmological constant. Studies of anisotropies in the Cosmic Microwave Background (CMB) with experiments like Planck \citep{Planck18} have at the same time provided percent-level constraints on other key cosmological parameters. Another more recent probe of cosmic expansion history is the measurement of Baryon Acoustic Oscillations (BAO), first in galaxy clustering \citep{Cole2005,EisensteinBAO} and later with the Lyman-$\alpha$ (\lya) forest \citep{Busca2013,B17,dmdb2020}. In each case the observable is a feature imprinted in the clustering measurements at comoving separation $\sim$100 $h^{-1}$Mpc. This feature was created at the epoch of recombination, where sound waves that were propagating through the dense photon-baryon plasma ceased to travel because of proton-electron recombination. The overdense regions at wave peaks gave rise to the preferential clustering scale we observe as BAO today. This scale can be measured in the CMB \citep{Planck18}, and as such can be used as a standard ruler to measure the expansion of the universe \citep{DESIw0wa}.


The Dark Energy Spectroscopic Instrument (DESI) \citep{snowmass2013,desi_pt1,DESI_inst,DESI_overview,DESI_corrector,DESI_fiber_system,JGuySpectro} survey will measure the spectra and redshifts of more than 40 million galaxies \citep{HahnBGS,RaichoorELG,ZhouLRG,DESI_survey_ops} and 3 million quasars \citep{Chaussidon_2023} over the course of a 5-year survey, covering 14\,000 deg$^2$ of the sky. Precise measurements of large-scale structure have already been made in galaxy and quasar clustering for the first year of observations of DESI \citep{DESIGalaxies,DESI_full_shape_y1}, comprising spectra of targets between redshifts $0.1 < z < 2.1$, and the \lya forest at $z > 2.1$ \citep{DESILYA}. The \lya forest is a series of absorption lines present in the spectra measured from very distant quasars. They are created when light from these quasars intercepts neutral hydrogen in the intergalactic medium (IGM), and as such, are an effective tracer of large-scale structure. A BAO measurement from the \lya forest, made using a combination of the autocorrelation of the \lya flux-transmission field and its cross-correlation with quasars \citep{Font_XCF_BAO_2014,FontXCF}, has a higher effective redshift than DESI galaxies (2.33 \citep{DESILYA}). This allows us to probe an earlier stage of the Universe than the DESI galaxy sample, which additionally helps to break degeneracies in cosmological constraints \citep{DESIw0wa}. 


It is possible to drastically improve the cosmological constraining power of the \lya forest (and galaxy) clustering by including information from the Alcock-Paczynski (AP) effect \citep{CuceuAPMethod, CuceuAPeBOSS}. It was projected in \cite{CuceuAPMethod} that including AP information from both the BAO peak and the smooth component of the \lya clustering measurement will give 2.5$\%$ and 1$\%$ constraints on $\rm \Omega_m$ and $H_{\rm 0}r_{\rm d}$, from the full DESI survey. This represents a factor of $\sim$2 improvement over BAO-only analyses. Additionally, a sub-10$\%$ level measurement of $f\sigma_8$ (at $z \sim 2.33$) should be possible when combining AP + BAO measurements from the full set of \lya correlation functions with redshift-space distortion (RSD) constraints derived from the quasar autocorrelation function.

It was demonstrated in \cite{DESILYA} that the analysis method used to measure BAO is quite robust to various systematic effects, in large part due to the decomposition of the peak and smooth components of the correlation function \citep{Kirkby2013}. However, since we are interested in constraining both the peak and smooth components of the AP effect, we expect to be more sensitive to contaminants than the BAO-only measurement. 
The main contaminants usually considered are continuum fitting distortion \citep{slosar_boss_y1}, high column density (HCD) systems \citep{FontHCD2012,RogersHCD2018,IgnasiSBLA}, broad absorption line (BAL) features \citep{BALImpact}, metal absorption \citep{FontHCD2012,PieriCGM} and redshift errors in quasars arising from both astrophysical peculiar velocities and measurements \cite{FontXCF}. Both HCDs and BALs are identified using neural-network--based finder algorithms \citep{BALFinder,DLAFinder} and masked to reduce their impact on the correlation functions. The former is also modelled to account for bias introduced by lower-density, undetected systems. Metals are modelled successfully by treating each absorber as a biased tracer of structure, where the biases are marginalised during the BAO fits.

Random velocities of galaxies or quasars distort clustering measurements along the line of sight, producing anisotropies relative to the transverse direction — an effect known as Fingers of God (FoGs). This effect will be present in the cross-correlation of \lya forests with quasars \citep{FontXCF}, which we will marginalise over using a version of the model from \cite{Percival09} (see section \ref{sec:method}). In the \lya autocorrelation, since the redshifts of \lya absorbers are set by intergalactic hydrogen and not by quasar motion, we do not need to model this effect. 

Errors in redshift measurement produce both the FoG effect and a second, distinct effect on our correlation functions. When computing the \lya transmission field $\delta$, we use a weighted mean continuum of all forests (equations \ref{eq:delta_main}, \ref{eq:meancontdelta}). As first described in \cite{Youles}, this mean continuum is smoothed by quasar redshift errors, causing spurious features in our correlation functions. They also presented a model for this contamination in the cross-correlation function, and showed that it had a small impact on the accuracy of their BAO constraints. This was also confirmed as a source of systematic bias in DESI DR1 \citep{BAO_VALIDATION_24} and in DESI DR2 \citep{Casas25}. In this paper we extend this work by modelling the same contamination in the \lya autocorrelation, and introducing free parameters that we marginalise over in our full-shape analysis.
 
To measure redshifts of DESI quasar spectra, a package called \texttt{Redrock} is used \citep{Redrock}. It is a template-fitting algorithm that uses Principal Component Analysis (PCA)-based quasar templates to determine the best-fit redshift in a chi-squared ($\chi^2$) minimisation. We expect a degree of systematic error in these estimations due to spectral variation and limited performance of templates, which was found to be $\sim340\,$km/s \citep{DESI_QSO_VI} for DESI survey validation. These variations are caused by Doppler shift of quasar emission lines due to the physical processes happening in the line-emitting region \citep{Shen16}. The typical level of shift away from the systemic quasar redshift depends on the line, but highly ionised, broad emission lines tend to exhibit the largest shifts. Recent work on new templates have significantly reduced this error for DESI Data Release 1 (DR1; \cite{DR1}), showing it to be $\sim$50\% smaller than survey validation \citep{Bault24,AllysonTemplates}. This was partly done by incorporating precise measurements of the Mg\,II line, which has relatively low bias with respect to the systemic redshift of the quasar. Broad emission lines within the \lya forest, listed in table \ref{tab:elines} \citep{BOSSComposite}, can also be Doppler-shifted, potentially contributing to errors in the mean continuum estimation (see section \ref{sec:cont_fitting}).



This paper is organised as follows. In section \ref{sec:method}, we describe the mock datasets used to develop and test our model, the method for generating the flux transmission field from quasar spectra, and the procedures for computing and modelling our correlation functions. In section \ref{sec:ImpactofErrors}, we show measurements of the \lya auto- and \lya–quasar cross-correlations from different mock datasets, both with and without redshift errors. Section \ref{sec:model} presents the redshift error model for both the auto- and cross-correlation functions, and section \ref{sec:fits} shows results from AP and isotropic BAO fits on our mock datasets. In section \ref{sec:fits_to_data}, we discuss the application of our model to real data and how the contamination can be mitigated using data cuts. Finally, section \ref{sec:summary} summarises our findings. 






\section{Method}\label{sec:method}
In this section we present the mock datasets we use (section \ref{sec:datasets}), and the analysis techniques employed to derive measurements of BAO and AP parameters from synthetic quasar spectra. The latter part begins with the flux transmission field measurement in section \ref{sec:cont_fitting}, followed by a description of how our correlation functions are constructed in section \ref{sec:corrfun}. Finally, we describe the model of the correlation functions and how we constrain BAO and AP parameters in sections \ref{sec:BAOAP} and \ref{sec:corrmod}.

\subsection{Data sets}\label{sec:datasets}

To test our model, we use synthetic data sets developed to validate the \lya BAO analysis in \cite{DESILYA}, described in detail in \cite{BAO_VALIDATION_24,quickquasars}. These are based on the \texttt{CoLoRe}\footnote{https://github.com/damonge/CoLoRe} \citep{CoLore} suite, which generates log-normal density fields with quasars distributed according to a specific biasing model. Lines-of-sight (also called skewers) are drawn from quasar positions in this field, with modified small-scale power, and converted to transmitted flux fraction using \texttt{LyaCoLore}\footnote{https://github.com/igmhub/LyaCoLoRe} \citep{LyaCoLore}. The clustering of these noiseless skewers is designed to be realistic at scales relevant for BAO systematics studies ($\sim$100 $h^{-1}$Mpc). \texttt{LyaCoLore} also stores skewers of metal absorption and can include high column-density (HCD) systems, both of which are major contaminants in standard \lya BAO analyses. 

To generate realistic quasar spectra from the flux transmission skewers, we use a final package called \texttt{desisim}\footnote{\url{https://github.com/desihub/desisim}} \citep{quickquasars}. This takes the transmission skewers of \texttt{LyaCoLore} as an input, multiplies them by a quasar continuum template, and adds noise to mimic the observing conditions and the instrumental model of DESI. Continuum templates are generated using \texttt{SIMQSO}\footnote{https://github.com/imcgreer/simqso} \citep{SIMQSO}, which combines a broken power law with a series of Gaussian emission lines. The slopes of the power law are sampled from a Gaussian,  with mean and dispersion tuned to better reflect the continuum shape and variability of quasars in the eBOSS DR16 dataset \citep{dmdb2020}. The \lya forest emission lines are simulated from the composite model of BOSS spectra \citep{BOSSComposite}, with line diversity is drawn from the distribution of equivalent widths (EWs). In table \ref{tab:elines}, we show the properties of these emission lines.

\begin{table}
    \centering
    \begin{tabular}{c|c|c|c}
    \hline
       Line  &  Wavelength [\AA]  & Equivalent Width [\AA] & $\sigma$ [\AA]\\
    \hline
       SiIV & 1064/1074 & 2.9/0.7 & 7.7/3.5 \\
        NII  & 1083 & 1.32 & 5.3 \\
         PV  & 1118/1128 & 0.76/0.46 & 5.3/4.1 \\
         CIII*  & 1175 & 2.49 & 7.7 \\

    \end{tabular}
    \caption{The emission lines within the \lya forest included in our synthetic spectra \citep{BOSSComposite}. We show from left to right, their rest-frame wavelength (in Angstroms), the mean equivalent widths and mean FWHM.}
    \label{tab:elines}
\end{table}

To emulate the effect of redshift errors in our mock datasets, we add Gaussian-distributed velocities $dv$ with zero mean and a dispersion of 400 km/s to each quasar redshift ($1+{z}_{q} = (1 + z_{q}^0)(1+dv/c)$). The reason for doing this rather than using the error from a redshift-fitting algorithm is that the latter has been shown to perform better on simulated spectra than real data \citep{Farr2020}. In \cite{Youles}, they differentiate between the Fingers of God (FoG) effect and redshift errors that affect the mean continuum. In this paper we focus on the latter.

We use two synthetic datasets: one has redshift errors added only to the tracer quasar catalogue, emulating the effect of FoGs. In the other we add redshift errors immediately after quasar spectra are generated, which will additionally give rise to mean continuum errors (see section \ref{sec:model}). Note that in our measured correlation functions, the latter set will exhibit the combined effect of FoGs and mean continuum errors. This means that later in section \ref{sec:ImpactofErrors} when we compare the differences between the two sets, we will observe only the contamination caused by continuum errors.

Each set consists of a stack of 100 mocks designed to emulate the survey conditions of DESI DR1 \citep{DR1}, including target density, footprint and signal-to-noise ratio. We will refer to these as DR1 mocks going forward. They also include all of the major contaminants affecting the analysis on data: HCDs, metal absorption and broad absorption lines (BALs). Using this synthetic dataset, we can validate our model and test correlations between various parameters in a controlled environment.






\subsection{Continuum fitting}\label{sec:cont_fitting}

The flux transmission field used to perform our clustering analysis is defined as:

\begin{equation}\label{eq:delta_main}
   1 + \delta(\lambda) = \frac{f(\lambda)}{\overline{F}(\lambda)C_{q}(\lambda)} ,
\end{equation}

\noindent where $f(\lambda)$ is observed flux, $\overline{F}(z)$ is the mean transmission and $C_{q}(\lambda)$ is the unabsorbed quasar continuum. Our analysis is discretised into "pixels" at the dispersion of the DESI spectrograph, 0.8\AA. At large enough scales, the delta field is a linear tracer of the underlying dark matter, but at small-scales the relationship becomes more complex due to non-linear structure formation and gas dynamics. Since we can't measure the unabsorbed continuum $C_{q}(\lambda)$ directly, we re-characterise its product with the mean transmission $\overline{F}(z)$:

\begin{equation}\label{eq:meancontdelta}
   \overline{F}(\lambda)C_{q}(\lambda) = \overline{C}(\lambda^{\rm rf}) \left( a_{q} + b_{q}\frac{\log\lambda - \log\lambda_{\rm min}}{\log\lambda_{\rm max} - \log\lambda_{\rm min}} \right).
\end{equation}

\noindent The term $\overline{C}(\lambda^{\rm rf})$, referred to as the "mean continuum", is the weighted mean of all forests in our analysis. This is multiplied by a first degree polynomial in $\log\lambda$ for each quasar $q$ in our sample, to account for continuum variability. The free parameters $a_{q}$ and $b_{q}$ are estimated by "continuum fitting", a process that involves maximising the log-likelihood:

\begin{equation}\label{eq:contfit_L}
    2\ln L = -\sum_{q} \frac{(f(\lambda) - \overline{F}(z)C_{q}(\lambda)(\lambda,a_{q},b_{q})}{\sigma_{q}^2(\lambda)} - \sum_{q} \sigma_{q}^2(\lambda) ,
\end{equation}

\noindent where $\sigma_{\rm q}^2(\lambda)$ is the flux variance, a combination of pipeline noise and large-scale structure. The mean continuum $\overline{C}(\lambda^{\rm rf})$ is also computed during this maximisation. For further details on exactly how this is done, see \cite{CesarLya}.

A notable side effect of fitting ($a_{\rm q}, b_{\rm q}$) using data from entire forests is that the measured $\delta_{q}(\lambda)$ will be distorted. In other words, our measured $\delta(\lambda_{i})$ in pixel $i$ will be a linear combination of all true underlying $\delta(\lambda_{j})$, such that $\delta_{i}' = \eta_{ij} \delta_{j}.$ The solution to this, described in detail in \cite{dmdb17,B17,DESILYA}, is to apply a linear projection operator to our measured $\delta'$ that explicitly removes the mean and slope of each quasar forest from the field:

\begin{equation}\label{eq:projdel}
    \Tilde{\delta}(\lambda) = \delta'(\lambda) - \sum_{j} \alpha_{qij}\delta_{q,j} , 
\end{equation}

\noindent where

\begin{equation}\label{eq:projdelmat}
    \alpha_{qij} = \frac{w_{q,j}}{\sum_{k} w_{q,k}} + \frac{w_{q,j} \Lambda_{\rm q,i} \Lambda_{q,j}}{\sum_{k} w_{q,k} \Lambda^2_{q,k}}.
\end{equation}

\noindent Here, $\Lambda_{i} = \log(\lambda_{i}) - \overline{\log \lambda_{i}}$, where $i$,$j$ and $k$ represent forest pixels. The reason for doing this is that the projected field $\Tilde{\delta}$ has the same distortion as our projected true field. We measure the correlations of our projected field $\Tilde{\delta}$ and, as we will show in section \ref{sec:corrmod}, project the model correlation function in the same way.

\subsection{Correlation functions}\label{sec:corrfun}
In this section we will show how we measure the \lya auto-correlation and the \lya-quasar cross-correlations from the projected transmission field outlined in the previous section. Firstly, we measure our correlations in configuration space as a function of transverse and line-of-sight separation $(r_\perp,r_\parallel)$. To convert from redshift and angular separation we do the following:

\begin{align}
    r_\perp & = (d_{\rm m}(z_{i}) + d_{\rm m}(z_{j})) \, \sin(\theta_{ij}/2) , \\ 
    r_\parallel &= (d_{\rm c}(z_{i}) - d_{\rm c}(z_{j})) \, \cos(\theta_{ij}/2) ,
\end{align}

\noindent where $d_{\rm m}$ is the transverse comoving distance and $d_{\rm c}$ is the comoving distance. A fiducial cosmology is required to compute both distances in terms of redshift, so we use the same Planck 2018 (hereby \texttt{P18}; \cite{Planck18}) cosmology as \cite{DESILYA,GordonEDR}. The correlation within each bin $(r_\perp,r_\parallel)\in\,A$ is then determined by a simple weighted average:

\begin{equation}\label{eq:corr_estimator}
    \xi_{A} = \frac{\sum_{i,j} w_{i} w_{j} \delta_{i} \delta_{j}}{\sum_{i,j} w_{i} w_{j}} ,
\end{equation}

\noindent where the weights are defined as:

\begin{equation}\label{eq:weighting}
    w_{i} = \frac{1}{\sigma^2_{i}} \left(1+z_{i}\right)^{\kappa - 1}.
\end{equation}

\noindent Here $\sigma_{i}$ is the variance from equation \ref{eq:contfit_L}, but with an extra term to increase the large-scale structure contribution. This term was implemented in \cite{CesarLya}, to maximise the signal-to-noise of our correlation functions. The second term in equation \ref{eq:weighting} takes into account the correlation function amplitude scaling with redshift. For \lya pixels, $\kappa = 2.9$ and for quasars $\kappa = 1.44$. Note that for the cross-correlation, index $j$ in equation \ref{eq:corr_estimator} represents a quasar, which we treat as points that effectively have $\delta_{j}=1$. This is equivalent to computing the weighted mean of $\delta_{\alpha}$ at set distances from a quasar. In both cases we choose a bin size of 4 $h^{-1}$Mpc following \cite{DESILYA,GordonEDR}. In the autocorrelation we measure between 0 and 200 $h^{-1}$Mpc in both directions, giving a total of 2500 bins. The cross-correlation however is not symmetric along the line-of-sight under permutation of pixels and quasars. Therefore, we define $r_\parallel$ between -200 and 200 $h^{-1}$Mpc, where negative separations correspond to a quasar being behind the \lya pixel with respect to the observer and vice-versa for positive separations. This gives us 5000 bins in total for the cross-correlation.

The covariance of our measurement, explained in detail in \cite{dmdb2020}, is estimated by splitting our dataset into sub-samples and computing:

\begin{equation}
    C_{AB} = \frac{\sum_{s} W^{s}_{A} W^{s}_{B} (\xi^{s}_{A} \xi^{s}_{B} - \xi_{A} \xi_{B})}{W_{A} W_{B}} ,
\end{equation}

\noindent where $A$ and $B$ are bins of our correlation function, and $s$ is a sub-sample. In our case, each \lya forest or quasar is uniquely set in one \texttt{HEALpix} sample $s$. The covariance is then smoothed by averaging all off-diagonal elements of the correlation matrix ($C_{AB}/\sqrt{(C_{AA}C_{BB}})$) with the same $\Delta r_\perp = r_\perp^{A} - r_\perp^{B}$ and $\Delta r_\parallel = r_\parallel^{A} - r_\parallel^{B}$. Furthermore, as introduced in \cite{DESILYA}, we now include the cross-covariance between the two correlation functions when doing fits to the model. 

In \cite{DESILYA} and \cite{LyaFS} they use two additional correlation functions that measure the autocorrelation of \lya absorption in a bluer region of the quasar ("region B", between 920-1020\,\AA \cite{LYADR2}) and its cross-correlation with quasars. For simplicity, and because the relative contribution to cosmological constraining power is small ($\sim10$\%), we choose not to include them in our analysis.

\subsection{BAO and full-shape information}\label{sec:BAOAP}
In this section we outline the model used to extract cosmological information and capture behaviour of contaminants in our analysis. The full-shape analysis that we outline here was first described for the \lya forest in \cite{CuceuAPMethod}, and performed on eBOSS data in \cite{CuceuAPeBOSS}. We follow the method of \cite{Kirkby2013} and employ a template power spectrum, based on \texttt{P18} cosmology, that we allow to vary from the measured cosmology by re-scaling our co-ordinate grid:

\begin{align}
    r_\parallel' = q_\parallel r_\parallel, & \hspace{+0.5cm}
    r_\perp' = q_\perp r_\perp.
\end{align}

\noindent We wish to distinguish between isotropic re-scaling $\alpha$ of the correlation function and anisotropic re-scaling $\phi$ (AP effect), therefore we define: 


\begin{align}
    \alpha(z) = \sqrt{q_\parallel q_\perp} \hspace{0.2cm} \rm{and} \hspace{0.2cm} 
    \phi(z) = \frac{q_\perp}{q_\parallel}
\end{align}

\noindent The first step in standard \lya BAO analyses is to take the linear isotropic input power spectrum from \texttt{P18}, $P_{\rm L}(k)$, and decompose it into peak ($P^{\rm peak}_{\rm L}(k)$) and smooth ($P^{\rm sm}_{\rm L}(k)$) components following \cite{Kirkby2013}. This is possible because the BAO is a distinct feature in the full correlation function, which also makes the standard peak-only analysis more robust to systematics affecting the smooth component. In the full-shape analysis we keep this peak-smooth decomposition, but introduce scaling parameters $\alpha$ and $\phi$ for each component:

\begin{equation}
    \xi(r_\perp,r_\parallel) = \xi_{\rm p}(r_\perp,r_\parallel,\alpha_{\rm p},\phi_{\rm p}) + \xi_{\rm s}(r_\perp,r_\parallel,\alpha_{\rm s},\phi_{\rm s}) ,
\end{equation}

\noindent where "s" and "p" refer to the smooth and peak components respectively. The peak component is defined in terms of distances and the sound horizon scale $r_{\rm d}$ as:


\begin{equation}\label{eq:BAOiso}
    \alpha_{\rm p} = \sqrt{\frac{d_{\rm m}(z)d_{h}(z)/r_{\rm d}^2}{[d_{\rm m}(z)d_{h}(z)/r_{\rm d}^2]_{\rm fid}}}
\end{equation}

\noindent where $d_{h}$ is the Hubble distance and "fid" is the fiducial \texttt{P18} cosmology. From this one can constrain the combination of $H_{0}r_{\rm d}$ and $\Omega_{\rm m}$, where $H_{0}$ is the Hubble constant. Both AP components are equivalent to $\phi=d_{\rm m}(z)H(z)/[d_{\rm m}(z)H(z)]_{\rm fid}$. From this we can directly measure $\Omega_{\rm m}$, and therefore break the degeneracy between $\Omega_{\rm m}$ and $H_0 r_{\rm d}$. A more detailed description of this is given in \cite{CuceuAPMethod}.

In \cite{CuceuAPMocks}, it was shown that including the smooth component AP parameter $\phi_{\rm s}$ doubled the precision of constraints on $\Omega_{\rm m}$. Note that in \cite{CuceuAPMocks} and in this work, we treat $\alpha_{\rm s}$ as a nuisance parameter since it is not clear how to extract cosmological information from it, and it's more strongly correlated with the \lya bias. Finally, while the information gain of including the smooth-component is substantial, it is also more susceptible to contaminants, thus motivating us to create the model presented in this paper.

\subsection{Correlation model}\label{sec:corrmod}

The \lya and \lya-quasar cross power spectra are given by linear perturbation theory, with a Kaiser \citep{K87} model of RSD:

\begin{align}
    & P_{\rm \alpha}(k, \mu_{\rm k}, z) = b_{\alpha}^2 (1 + \beta_{\alpha}\mu_{\rm k}^2)^2 F_{\rm sm} P_{\rm L}(k, z), \\
    & P_{\rm X}(k, \mu_{\rm k}, z) = b_{\alpha}b_{\rm q}(1+\beta_{\alpha}\mu_{\rm k}^2)(1+\beta_{q}\mu_{\rm k}^2) F_{\rm sm} F_{\rm NL,q} P_{\rm L}(k, z) ,
\end{align}

\noindent where $b_{i}$ are the cosmological biases and $\beta_{i}$ are the RSD parameters for tracer $i$ (quasar or \lya pixel), and $\mu_{\rm k}$ is the cosine of the k vector with respect to the line-of-sight direction. The cross-correlation term $F_{\rm NL,q}$ accounts for quasar peculiar velocities, and is given in \cite{Percival09}:


\begin{equation}
    F_{\rm NL,q}(\rm k_\parallel) = \sqrt{\frac{1}{1 + (\rm k_\parallel \sigma_{\rm v})^2}}.
\end{equation}

\noindent This produces a smoothing effect at large $\rm k_\parallel$, where $\sigma_{\rm v}$ is a free parameter in our fit. This is a distinct effect from the redshift error continuum smoothing we model in this paper. We also account for any systematic quasar redshift errors by introducing a free parameter to our cross-correlation model: $\Delta r_\parallel = r_{\parallel,{\rm true}} - r_{\parallel,{\rm measured}}$. 

$F_{\rm sm}$ is a Gaussian smoothing term that accounts for our log-normal simulations having limited grid size ($\sim$2.4 $h^{-1}$Mpc). This has two free parameters ($\sigma_\perp,\sigma_\parallel$) which we marginalise over in our analysis. 

\begin{figure*}
    \centering
    \includegraphics[width=1\linewidth]{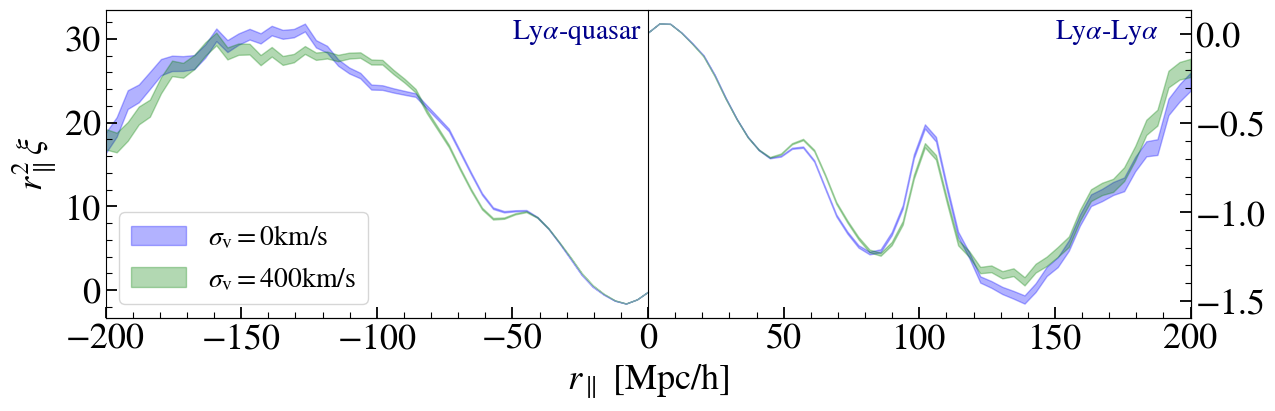}
    \caption{The effect of continuum redshift errors on the \lya-quasar cross-correlation (left) and the \lya-\lya auto-correlation (right) computed from our mock datasets. Each dataset is the stack of 100 DESI DR1 mocks. We take the weighted average over the first 4 bins in transverse separation $r_\perp$ ([0,16] $h^{-1}$Mpc) where the effect of redshift errors is strongest, and plot as function of line-of-sight separation $r_\parallel$. We only include negative $r_\parallel$ for the cross-correlation, since as shown in figure \ref{fig:2dzerr_auto} there is no impact at $r_\parallel>0$ $h^{-1}$Mpc. Note that at $\sim\pm60$ $h^{-1}$Mpc we see bumps in the correlations due to SiII(1190/1193) lines.}
    \label{fig:auto_cross_corr}
\end{figure*}

In the contaminated mock dataset we introduce metal absorption and HCD contamination. For HCDs, we mask systems with column density $>2 \times 10^{20}\rm cm^{-2}$, following the analysis of \cite{DESILYA}. The remaining HCDs, which in real data are too small to detect for masking, are biased tracers of large-scale structure. Following \cite{FontHCD2012}, we treat re-write the \lya bias and RSD parameters as effective combinations of signal from \lya forest and HCDs:

\begin{align}
    b_{\alpha}' & = b_{\alpha} + b_{\rm HCD} \ \rm exp \left( -L_{\rm HCD} \rm k_\parallel \right),\\
    b_{\alpha}'\beta_{\alpha} & = b_{\alpha}\beta_{\alpha} + b_{\rm HCD} \beta_{\rm HCD} \ \rm exp \left( -L_{\rm HCD} \rm k_\parallel \right),    
\end{align}

\noindent where $L_{\rm HCD}$ is the typical length-scale of the unmasked HCDs. In our fits we marginalise over this and the bias and RSD of the HCD systems. 

In our analysis, we assume all of the absorption within the \lya forest is due to the \lya transition. However, in some cases we are mistaking \lya absorption with absorption from different metals. This leads to a contamination which we correct for using the same process outlined in \cite{DESILYA}. In our fits, we marginalise over a set of bias parameters for 4 metal lines: SiIII(1207), SiII(1190), SiII(1193), and SiII(1260) (section \ref{sec:fits}).




As mentioned in section \ref{sec:cont_fitting}, distortions introduced to $\delta$ during continuum fitting lead us to use a "projected" field $\Tilde{\delta}$ (equation \ref{eq:projdel}). This is designed such that the projected distorted field has the same distortion as the projected true field. We write the distorted model ($A$) in terms of the undistorted model ($B$) using a "distortion matrix": $\xi_{A} = \sum_{AB} D_{AB} \xi_{B}$. $D$ can be written as:

\begin{equation}
    D_{AB} = \frac{1}{W_{A}}\sum_{ij \in A} w_{i}w_{j} \left(\sum_{i'j' \in B} \eta_{ii'} \eta_{jj'} \right) ,
\end{equation}

\noindent where $\eta_{mn} = \delta^{\rm K}_{mn} - \alpha_{mn}$. Here, $ \delta^{\rm K}$ is the Kronecker delta, and $\alpha$ is the projection matrix of equation \ref{eq:projdelmat}. Note for the cross-correlation $\eta_{jj'}=1$, since there is only one correlating forest. To compute the auto- and cross-correlation models in our analysis we use the package \texttt{Vega}\footnote{https://github.com/andreicuceu/vega}. This is also used to fit cosmological parameters, shown in section \ref{sec:fits}.

\section{Impact of redshift errors}\label{sec:ImpactofErrors}

In this section we will show the impact of continuum redshift errors on the measured correlation functions of our mock datasets. Then, in section \ref{sec:model}, we will discuss the origin of this contamination and propose a model for it.

We use the two synthetic datasets described in section \ref{sec:datasets}, both of which are stacks of 100 DESI DR1 mocks. The "uncontaminated" set has redshift errors added to the tracer quasar catalogue, emulating FoGs. The "contaminated" set has errors added to the spectra, emulating pipeline redshift errors. Using this method we isolate the effect of redshift errors which enter during the continuum fitting process (section \ref{sec:cont_fitting}). We refer to this contamination as continuum redshift errors, to differentiate it from the effect of FoGs.

We then compute the \lya autocorrelation and its cross-correlation with quasars using the method outlined in the previous section. For each of the two sets, we compute the weighted mean and covariance of the 100 correlation functions. We use these two stacks to model and fit the effect of continuum redshift errors in our correlation functions in the following sections. The auto- and cross-correlation functions from the uncontaminated and contaminated sets are shown in figure \ref{fig:auto_cross_corr}. These are plotted as a function of the line-of-sight separation, averaged over the first 4 bins in transverse separation ([0,16] $h^{-1}$Mpc) where the effect is strongest. We can also clearly see a peak at $\sim100$ $h^{-1}$Mpc, a combination of BAO and SiII(1260) contamination, and other peaks caused by metal contamination (section \ref{sec:corrmod}).

\begin{figure}
    \centering
    \begin{subfigure}{0.45\textwidth}
        \centering
        \includegraphics[width=\linewidth]{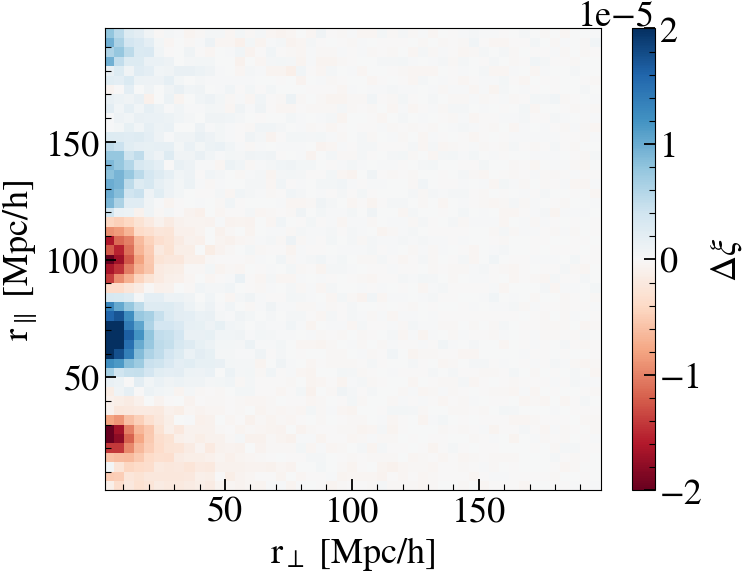} 
        \label{fig:2dzerr_auto}
    \end{subfigure}
    \hfill
    \begin{subfigure}{0.45\textwidth}
        \centering
        \includegraphics[width=\linewidth]{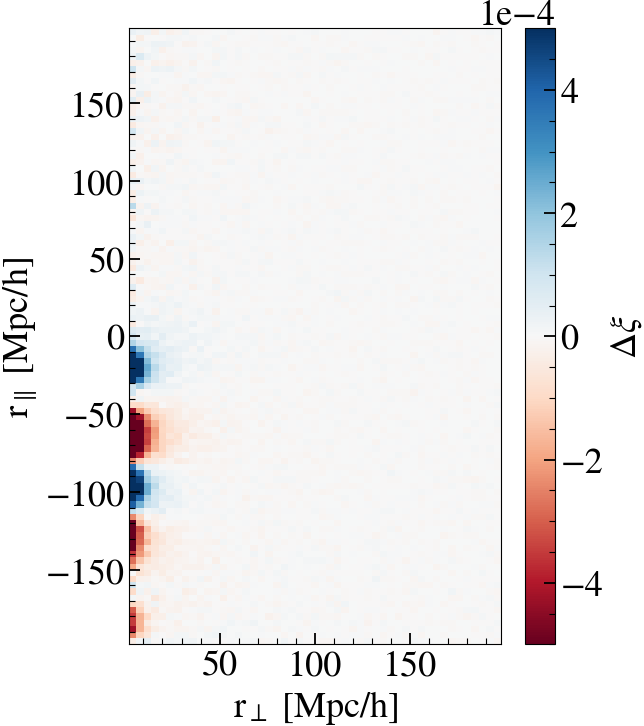} 
        \label{fig:2dzerr_cross}
    \end{subfigure}
    \caption{(Top) difference in the \lya auto-correlation measured from contaminated (with continuum redshift errors) and uncontaminated datasets, as a function of ($r_\perp,r_\parallel$). (Bottom) difference in the \lya-quasar cross-correlation from contaminated and uncontaminated datasets.}
    \label{fig:2dzerr_auto}
\end{figure}


The contamination of continuum redshift errors (the difference between the two datasets) in the auto- and cross-correlations is shown in figure \ref{fig:2dzerr_auto}. As discussed in \cite{Youles}, there is a strong dependence of redshift error distortion in the cross-correlation on the small-scale quasar autocorrelation, which is significantly weaker for positive $r_\parallel$ where host quasars of \lya pixels are much further from the correlating quasars. Therefore, as we see in figure \ref{fig:2dzerr_auto}, the distortion from redshift errors is only visible for negative $r_\parallel$. This dependence on the quasar autocorrelation is also the reason the contamination becomes negligible at transverse separations $r_\perp\gtrsim$20 $h^{-1}$Mpc. The behaviour is similar in the \lya autocorrelation, except the dependence is now on the small-scale \lya-quasar cross-correlation. In section \ref{sec:model}, this dependence can be seen explicitly in our model. The distortions in figure \ref{fig:2dzerr_auto} are also localised around specific values of $r_\parallel$. In section \ref{sec:evolwav} we show how this pattern arises due to smoothing of the mean continuum.

We can see that the continuum redshift errors will contaminate our full-shape measurement, but also that part of the contamination overlaps with the fiducial BAO position ($\sim$100 $h^{-1}$Mpc). In section \ref{sec:fits}, we show the effect of this contamination on the set of cosmological parameters measured from the BAO peak ($\alpha_{\rm p}$, $\phi_{\rm p}$) and the broadband component ($\phi_{\rm s}$, $f\sigma_8$) of our correlation functions. 




\section{Modelling the contamination}\label{sec:model}
In this section we will outline our model for continuum redshift error contamination. We start describing the effect of redshift errors on the mean continuum and flux-transmission field (following \cite{Youles}), and carry this through to a full model of the contamination in the autocorrelation. We also highlight the cross-correlation redshift error model of \cite{Youles}.




\begin{figure*}
    \centering
    \includegraphics[width=0.9\linewidth]{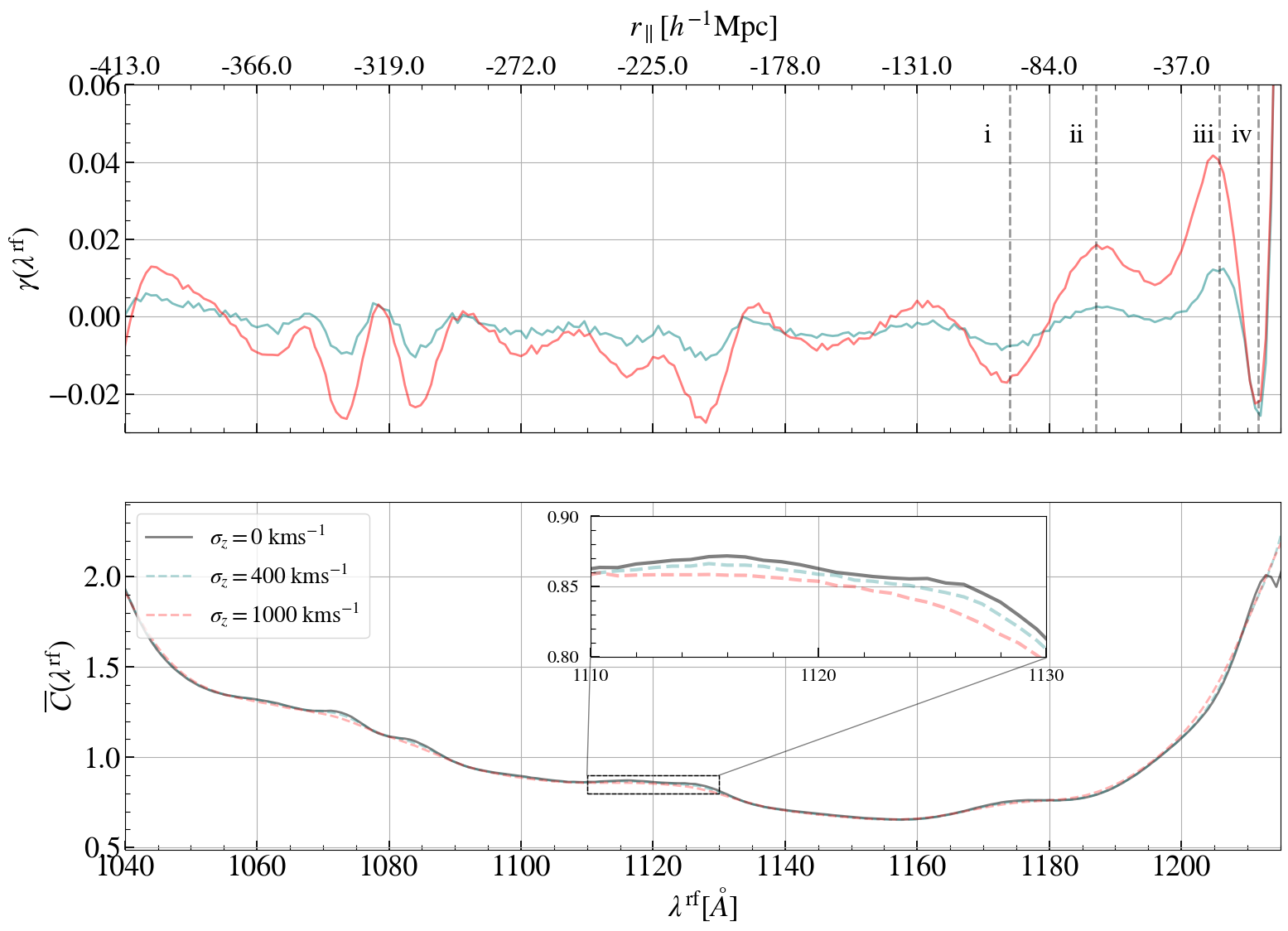}
    \caption{(Top) the mean continuum distortion function $\gamma=\hat{\overline{C}}/\overline{C} - 1$ as a function of rest-frame wavelength. $\hat{\overline{C}}(\lambda_{\rm rf}, \sigma_{\rm v})$ is the mean continuum with redshift errors $\sigma_{\rm v}$, as shown in the plot above. Roman numerals mark the location of prevalent features in $\gamma$, which contaminate our correlation functions. We include the approximate comoving distance between a quasar at $z = 2.3$ and pixels in its forest, along the top axis. (Bottom) the \lya forest mean continuum (equation \ref{eq:meancontdelta}) of our mock datasets with 0\,$\rm kms^{-1}$, 400\,$\rm kms^{-1}$ and 1000\,$\rm kms^{-1}$ of redshift errors added. The later is for visualisation while 400\,$\rm kms^{-1}$ is used in the actual analysis.}
    \label{fig:comb_cont_gamma}
\end{figure*}

\begin{figure}
    \centering
    \includegraphics[width=1\linewidth]{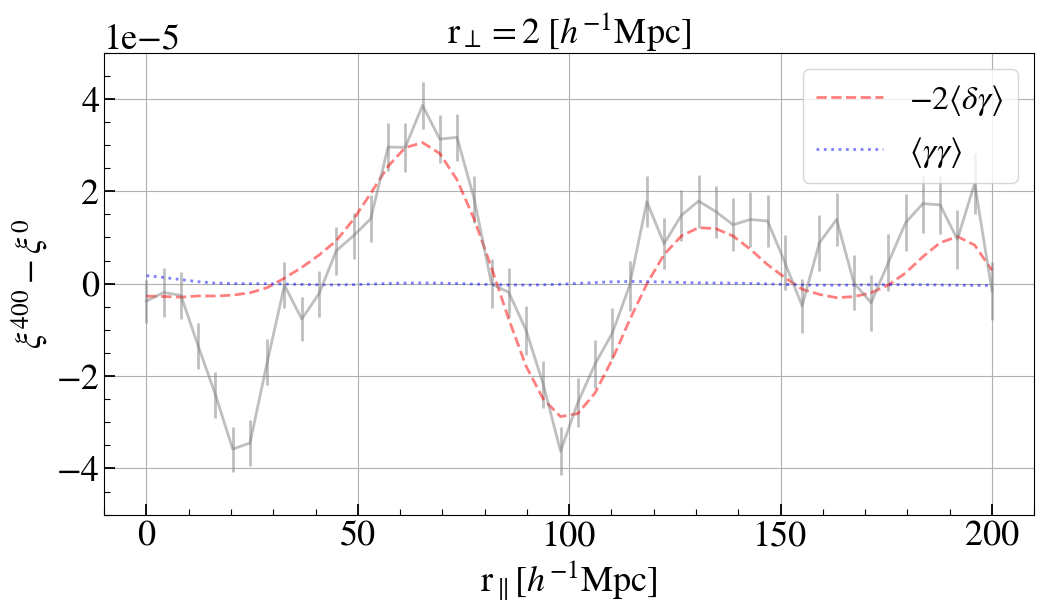}
    \caption{The difference in the \lya autocorrelation function between contaminated (with continuum redshift errors) and uncontaminated datasets (grey), overlaid with direct measurements of $\langle \delta \gamma \rangle$ (red dashed) and $\langle \gamma \gamma \rangle$ (blue dotted). We plot only the first $r_\perp$ bin where the contamination is strongest.}
    \label{fig:dg_measured}
\end{figure}

\begin{figure}
    \centering
    \includegraphics[width=1.1\linewidth]{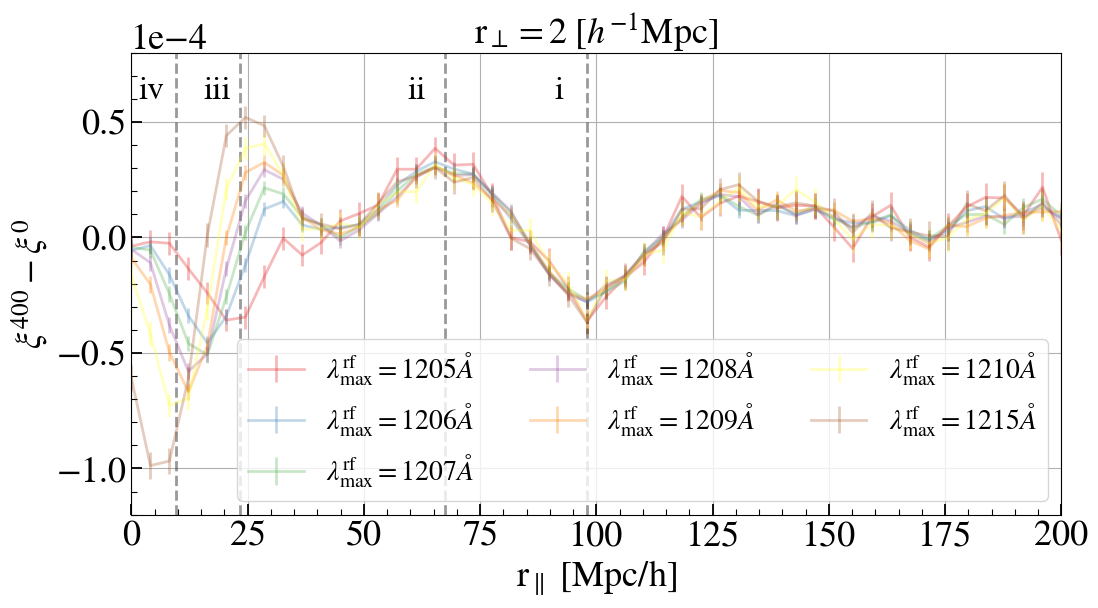}
    \caption{The evolution of continuum redshift error contamination with the maximum rest-frame wavelength of the \lya forest. The 1205\AA limit (red) is the limit used in the analysis of this paper. We include Roman numerals to indicate the features in $\gamma$ (figure \ref{fig:comb_cont_gamma}) which correspond to spurious correlations in this figure.}
    \label{fig:devol_rf}
\end{figure}

In \cite{Youles}, redshift errors were identified to smooth the mean continuum (equation \ref{eq:meancontdelta}) that is fit during the transmission field estimation. They subsequently defined a term $\gamma(\lambda^{\rm rf})$:

\begin{equation}\label{eq:gamma}
    \gamma(\lambda^{\rm rf}) = \frac{\hat{\overline{C}}(\lambda^{\rm rf})}{\overline{C}(\lambda^{\rm rf})} - 1 ,
\end{equation}

\noindent where $\hat{\overline{C}}$ is the mean continuum estimate with redshift errors and $\overline{C}$ is the true mean continuum. The \lya forest region of our mock continua consists of a power law smooth component, superimposed with many Gaussian emission lines. If quasar redshift errors are randomly distributed, the mean continuum will be smoothed around emission lines as shown in figure \ref{fig:comb_cont_gamma}. In appendix \ref{sec:appB}, we write an expanded version of $\gamma(\lambda^{\rm rf})$, accounting for the fact that the wavelength grid $\lambda^{\rm rf}$ for each forest is shifted from the true grid, due to errors in $z_{\rm q}$.

In the top panel of figure \ref{fig:comb_cont_gamma}, we show a measurement of $\gamma(\lambda^{\rm rf})$ formed using mean continua from our mock datasets with redshift errors of 400\,kms$^{-1}$ and 0\,kms$^{-1}$. Combining equations \ref{eq:gamma} and \ref{eq:delta_main} we get:

\begin{equation}
    \hat{\delta}(\lambda) = \frac{1 + \delta(\lambda)}{1 + \gamma(\lambda^{\rm rf})} - 1.
\end{equation}

\noindent Assuming we can ignore second-order terms:

\begin{equation}
   \hat{\delta}(\lambda) \approx \delta(\lambda) - \gamma(\lambda^{\rm rf}).
\end{equation}

\noindent Thus the measured autocorrelation function can be written as:

\begin{equation}\label{eq:dg}
   \langle \hat{\delta}(\lambda)\hat{\delta}(\lambda) \rangle = \langle\delta(\lambda)\delta(\lambda)\rangle - 2\langle \delta(\lambda)\gamma(\lambda^{\rm rf})\rangle +  \langle\gamma(\lambda^{\rm rf})\gamma(\lambda^{\rm rf})\rangle.
\end{equation}

\noindent To determine the most dominant term in this equation, we directly compute $\langle\delta\gamma\rangle$ and $\langle\gamma\gamma\rangle$ using $\delta$ from mock datasets with no added redshift errors, and the measured gamma in figure \ref{fig:comb_cont_gamma}. The result is shown in figure \ref{fig:dg_measured}, where both terms are plotted over the measured contamination in the autocorrelation. Clearly, $\langle\gamma\gamma\rangle$ is negligible. Therefore, we only model the contribution of the $\langle\delta\gamma\rangle$ term. Noticeably, the trough in the correlation difference at $\sim25\,h^{-1}$Mpc is not captured by either term. In the following section we discuss the behaviour of this feature.

\subsection{Variation with wavelength}\label{sec:evolwav}

Spurious features in $\langle\delta\gamma\rangle$ caused by redshift errors in the \lya continuum appear at specific separations. These separations (in comoving distance) are determined by the difference in rest-frame wavelength between the \lya line and strong features in $\gamma(\lambda^{\rm rf})$:

\begin{equation}
\label{eq:rfeature}
   r^{\rm feature} \approx \frac{c(1+\overline{z})}{H(\overline{z})} \frac{(\lambda^{\rm feature} - \lambda^{\alpha})}{\lambda^{\alpha}},
\end{equation}

\noindent where $\overline{z}$ is the effective redshift of the dataset and $H$ is the Hubble parameter. In the top panel of figure \ref{fig:comb_cont_gamma}, we highlight key features in Roman numerals: i at 1174\,\AA, ii at 1187\,\AA, iii at 1205.7\,\AA \ and iv at 1211.6\,\AA, the latter two of which fall outside of the fiducial rest-frame limits (1205\,\AA) of our clustering analysis. Using these values in equation \ref{eq:rfeature} with an effective redshift of 2.3 predicts spurious correlation features at separations of 99\,$h^{-1}$Mpc, 70\,$h^{-1}$Mpc, 26\,$h^{-1}$Mpc and 9\,$h^{-1}$Mpc respectively. In figure \ref{fig:devol_rf}, we show the contamination in the \lya autocorrelation as a function of the upper rest-frame wavelength limit of the forest. We also include Roman numerals here, indicating the aforementioned comoving separations, and see clearly that they align with the strongest spurious correlation features.

Looking specifically at the rest-frame limit of our analysis, 1205\,\AA (red line in figure \ref{fig:devol_rf}), we see a trough at $\sim25\,h^{-1}$Mpc. This roughly overlaps with the feature iii (at 1205.7\,\AA) in figure \ref{fig:comb_cont_gamma}, but is a trough where we expect a peak (peaks in $\gamma$ produce peaks in $\hat\xi_{\alpha\alpha}$). Therefore we posit that this trough is instead caused by feature iv in figure \ref{fig:comb_cont_gamma}, at a rest-frame wavelength of 1211.6\,\AA. This is possible because quasar redshift errors not only smooth the mean continuum, but also shift the rest-frame wavelength grid with respect to the truth. For any forest in our dataset, we apply the upper rest-frame wavelength limit by converting observed wavelength $\lambda$ into rest-frame, where $\lambda^{\rm rf} = \lambda / (1+\hat{z}_{\rm q})$ for a measured quasar redshift $\hat{z}_{\rm q}$. If $\hat{z}_{\rm q}>z_{\rm q}$, pixels which have $\lambda^{\rm rf}>1205$\AA \ will fall into the accepted \lya region. Consequently, values of $\gamma(\lambda^{\rm rf}>1205$\,\AA) will contaminate our analysis. 



\begin{figure}
    \centering
    \includegraphics[width=1\linewidth]{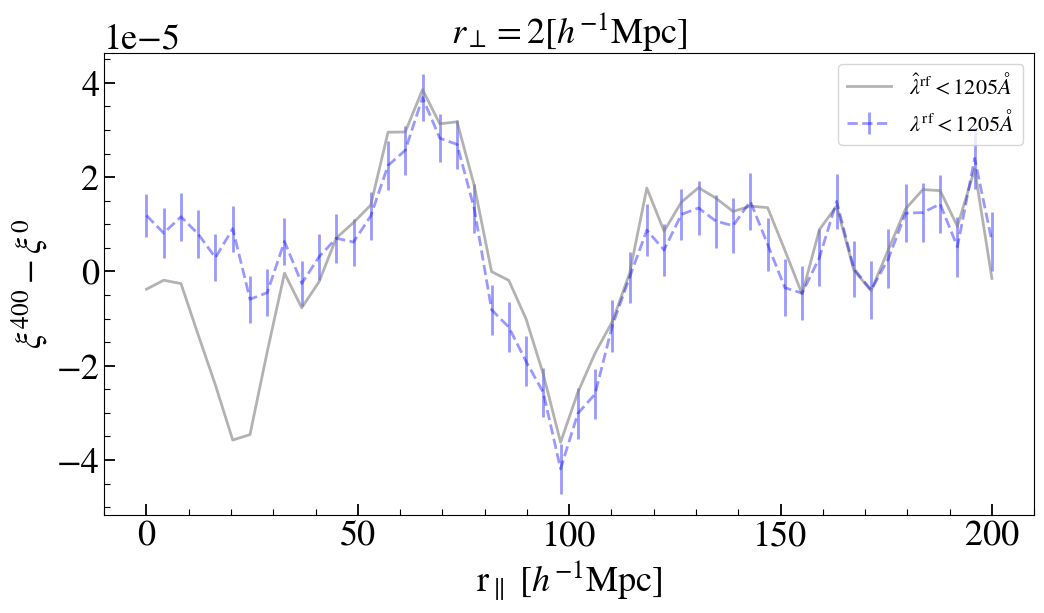}
    \caption{Contamination caused by redshift errors (grey), and the same contamination after removing pixels with true rest-frame wavelength greater than 1205\,\AA \ (blue dashed). After removing these pixels the trough feature at $\sim25\,h^{-1}$Mpc disappears, while the rest of the contamination is unaffected.}
    \label{fig:removing_pix}
\end{figure}

In figure \ref{fig:removing_pix}, we show (blue dashed line) the result of removing pixels (for each forest in our correlation measurement) which are in reality above the \lya rest-frame wavelength limit, but are accepted into the sample because of redshift errors. In this case, the trough feature is no longer visible, and the contamination at small-scales is now roughly consistent with 0. We have also verified that including pixels that are in reality below 1205\,\AA, but are excluded due to redshift errors, has no effect on the shape of the contamination.

In appendix \ref{sec:appB}, we make the shifting of the rest-frame wavelength grid where we evaluate each $\delta$ explicit, by expanding $\hat{\overline{C}}$ about small redshift deviations. This produces extra terms that contaminate the correlation functions, but to 1st order they do not capture the trough at 25\,$h^{-1}$Mpc in our fiducial analysis. We also try to empirically model the trough feature (equation \ref{eq:addcorr}), which works well in a direct fit (see figure \ref{fig:modfit}), but biases our full-shape analysis.

\subsection{Auto-correlation model}\label{sec:autocorr_mod}

To model the $\langle \delta(\lambda)\gamma(\lambda^{\rm rf})\rangle$ term in equation \ref{eq:dg}, we follow a similar approach to that used in \cite{Youles} to model the $\langle\gamma(\lambda^{\rm rf})\rangle$ term in the \lya-quasar cross-correlation model.

We begin by considering a general expression for the expectation value of $\langle \delta\gamma \rangle$:

\begin{equation}
    \langle\delta\gamma\rangle_{A} = \int_{-1}^{\infty} d\delta \int_{-1}^{\infty} d\gamma P(\delta,\gamma | \Vec{r}^A)\delta\gamma \,
\end{equation}

\noindent where $P(\delta,\gamma)$ is the probability of having a particular $\delta$ and $\gamma$, for separation $\Vec{r}^A=\{\Vec{r_\parallel}^A,\Vec{r_\perp}^A\}$ between two \lya pixels. This can be equivalently expressed as:

\begin{equation}\label{eq:dg_init}
    \langle\delta\gamma\rangle_{A} = \int_{-1}^{\infty} d\delta \int_{\lambda^{\rm rf,min}}^{\lambda^{\rm rf,max}} d\lambda^{\rm rf} P(\delta,\lambda^{\rm rf})\delta\gamma ,
\end{equation}

\noindent since $\gamma$ is uniquely defined by $\lambda^{\rm rf}$, and the integration limits are given by the fact that we only consider pixels within the defined \lya forest rest-frame region. We can write the quasar rest-frame wavelength of any pixel as:

\begin{equation}\label{eq:rf_to_zp}
    \lambda^{\rm rf} = \lambda^\alpha(1+z^{\alpha})/(1+z^{\rm Q}) ,
\end{equation}

\noindent where $\lambda^\alpha$ is the rest-frame wavelength of the \lya transition, and $z^{\alpha}$ is the redshift of a \lya pixel in the \lya forest region of a quasar with redshift $z^{\rm Q}$. Therefore, for a given $\lambda^{\rm rf}_{i}$ for forest $i$, we have a set of possible \{$z^{\rm Q}_{i}$, $z^{\alpha}_{i}$\}. Furthermore, for separations $r_\parallel \in A$ and pixel redshift $z^{\alpha}_{i}$, we have a unique $z^{\alpha}_{j}$ in forest $j$. Thus, we re-write equation \ref{eq:dg_init} as:
\begin{equation}\label{eq:dg_change_vars}
    \langle\delta\gamma\rangle_{A} = \int_{-1}^{\infty} d\delta_{j}  \int_{z^{\rm Q,min}}^{z^{\rm Q,max}} \int_{z^{\rm \alpha,min}}^{z^{\rm \alpha,max}} dz^{\rm Q}_{i} dz^{\alpha}_{j} P(\delta_{j}, z^{\rm \alpha}_{j},  z^{\rm Q}_{i} | r_\parallel^A) \delta_{j} \gamma_{i}.
\end{equation}

\noindent We can expand this further as:

\begin{multline}
    \label{eq:dg_bayes}
     \langle\delta\gamma\rangle_{A} = \int_{-1}^{\infty} d\delta_{j}  \int_{z^{\rm Q,min}}^{z^{\rm Q,max}} \int_{z^{\rm \alpha,min}}^{z^{\rm \alpha,max}} dz^{\rm Q}_{i} dz^{\alpha}_{j} P(\delta_{j} | z^{\rm \alpha}_{j},  z^{\rm Q}_{i}) \\ \times P(z^{\rm \alpha}_{j} |  z^{\rm Q}_{i}) P(z^{\rm Q}_{i}) \delta_{j} \gamma_{i} ,
\end{multline}

\noindent where we have made the dependence on $r_\parallel^A$ explicit. Expanding the first term in this equation, we get:

\begin{align}\label{eq:pexpansion}
    P(\delta_{j} | z^{\rm \alpha}_{j},  z^{\rm Q}_{i}) & = \frac{P(z^{\rm Q}_{i}|\delta_{j},z^{\rm \alpha}_{j}) P(\delta_{j} | z^{\rm \alpha}_{j})}{P(z^{\rm Q}_{i}|z^{\rm \alpha}_{j})}  \\
    & = \frac{P(z^{\rm Q}_{i}|\delta_{j},z^{\rm \alpha}_{j}) P(\delta_{j} | z^{\rm \alpha}_{j})P(z^{\rm \alpha}_{j})}{P(z^{\rm \alpha}_{j}|z^{\rm Q}_{i})P(z^{\rm Q}_{i})}.
\end{align}

\noindent Now we will make some approximations to evaluate the first two terms in the numerator of this equation - note that the first term in the denominator cancels with the second term in equation \ref{eq:dg_bayes}. We begin by writing $P(\delta_{j} | z^{\rm \alpha}_{j}) \approx P(\delta_{j})$. This is equivalent to ignoring the redshift evolution of $\delta$, normally proportional to $(1+z_{j}^{\alpha})^\kappa$, which we find does not change the shape of $\langle\delta\gamma\rangle_{A}$. 

Next, we express the first term in equation \ref{eq:pexpansion} as $P(z^{\rm Q}_{i}|\delta_{j},z^{\rm \alpha}_{j}) = P(z^{\rm Q}_{i})(1 + \delta_{j}\xi^{\rm X}(\Vec{r}^{\rm X}))$, where $\Vec{r}^{\rm X}$ is the separation between $z_{j}^\alpha$ and $z^{\rm Q}_{i}$, and $\xi^{\rm X}$ is the \lya-quasar cross-correlation. Note that the dependence of $P(z^{\rm Q}_{i}|\delta_{j},z^{\rm \alpha}_{j})$ on $z^{\rm \alpha}_{j}$ is accounted for by the redshift evolution of the cross-correlation. This is not an analytically derived expression, but rather an ansatz that produces the expected behaviour of $P(z^{\rm Q}_{i}|\delta_{j},z^{\rm \alpha}_{j})$. The definition of $\xi^{\rm X}$ here is somewhat analogous to the definition of the quasar autocorrelation as a measure of excess probability, $\xi^{\rm Q} = P(z^{\rm Q}_{j} | z^{\rm Q}_{i})/P(z^{\rm Q}_{j}) - 1$, but with a modulating $\delta$ term. Making the subsequent substitutions into equation \ref{eq:dg_bayes} gives:


\begin{multline}
    \langle\delta\gamma\rangle_{A} = \int_{-1}^{\infty} d\delta_{j} \int_{z^{\rm Q,min}}^{z^{\rm Q,max}} \int_{z^{\rm \alpha,min}}^{z^{\rm \alpha,max}} dz^{\rm Q}_{i} dz^{\alpha}_{j} P(\delta_{j}) P(z^{\rm \alpha}_{j}) P(z^{\rm X}_{i}) \\ \times (1 + \delta_{j}\xi^{\rm X}(\Vec{r}^{\rm X})) \hspace{+0.05cm} \delta_{j} \gamma_{i} .
\end{multline}

\noindent Multiplying the terms in brackets results in two parts (with and without the cross-correlation). The integration over $\delta$ is independent of the other two variables, and since it has mean 0, the part with one $\delta_{j}$ term vanishes. In the second term, $\int^{\infty}_{-1} P(\delta_{j})\delta_{j}^2 d\delta_{j}$ evaluates to $\sim 0.6$. We introduce a free amplitude for this model $A_{\rm cont}^{\rm auto}$, which will absorb this value, but here we call this factor $a_\delta$, and write our final expression as:

\begin{equation}\label{eq:dg_final}
    \langle\delta\gamma\rangle_{A} = a_\delta A_{\rm cont}^{\rm auto} \int_{z^{\rm Q,min}}^{z^{\rm Q,max}} \int_{z^{\rm \alpha,min}}^{z^{\rm \alpha,max}} dz_{i}^{\rm Q} dz_{j}^{\rm \alpha} \ P(z^{\rm \alpha}_{j}) P(z^{\rm Q}_{i}) \xi^{\rm X}(\Vec{r}^{\rm X}) \gamma_{i}.
\end{equation}

\noindent Qualitatively, we evaluate this model by iterating over correlation function bins ($r_\perp,r_\parallel$). For a bin $A$ (width 4\,$h^{-1}$Mpc in our analysis), we have N bins of $z^{\rm Q}_{i}$ covering the redshift range of the quasars in the dataset. This results in an array of N$\times$M pixel redshifts $z^\alpha_{i}$ (equation \ref{eq:rf_to_zp}), using M bins of rest-frame wavelength in the \lya forest region ($\lambda^{\rm rf}\in[1040,1205]$\r{A}). With these values we can compute $P(z^{\rm Q}_{i})$ from our normalised quasar redshift distribution, and the $\gamma_{i}$ term for each rest-frame wavelength in M. Then, given $z^\alpha_{i}$ and $r_\parallel \in A$, we can derive $z^\alpha_{j}$ and the vector $\Vec{r}^{\rm X}$ between $z^\alpha_{j}$ and $z^{\rm Q}_{i}$. Finally, we compute the probability of $z^\alpha_{j}$ using a normalised pixel redshift distribution, and evaluate the cross-correlation $\xi(\Vec{r}^{\rm X})$. 

\begin{figure}
    \centering    \includegraphics[width=\linewidth]{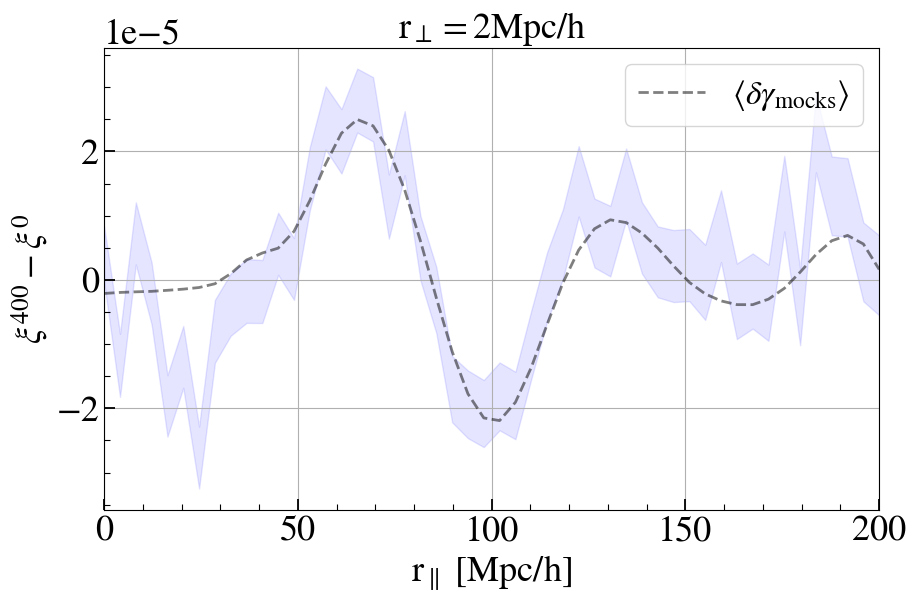}
    \caption{The contamination introduced by continuum redshift errors (blue shaded) as a function of $r_\parallel$, for the first (most contaminated) $r_\perp$ bin. The black dashed line is the contamination model (equation \ref{eq:dg_final}), estimated using a measurement of $\gamma(\lambda^{\rm rf})$.}
    \label{fig:gamma_mod_meas}
\end{figure}

In figure \ref{fig:gamma_mod_meas}, we present the model compared to the difference of correlation functions with and without redshift errors. To compute this, we directly measure $\gamma(\lambda)$ from the mock datasets (via equation \ref{eq:gamma}), and input it to equation \ref{eq:dg_final}. In theory, one could also use a measurement of $\gamma$ from mock datasets to fit our model on data. However, this relies on knowing the amount of redshift error present in the data, so we opt instead for building a simple model described in the next section.


\subsection{Modelling $\gamma$}\label{sec:modgamma}

In real data we cannot directly measure $\gamma(\lambda^{\rm rf})$, since we don't know the true continuum (without redshift errors). Therefore, we need to construct a model of $\gamma(\lambda^{\rm rf})$. 

Recalling equation \ref{eq:gamma}, we see $\gamma$ contains the ratio of the mean continuum in the presence of redshift errors ($\hat{\overline{C}}$), to the true mean continuum ($\overline{C}$). Therefore, to construct a model of $\gamma$, we need a template quasar continuum and a parameter that controls the amount of redshift error to input. The continuum template is a convolution of a set of emission lines and a smooth broadband component. For the former we use the emission line properties of the composite model of BOSS spectra DR9 \citep{BOSSComposite}. We then approximate the smooth component as flat, motivated by the fact that the majority of the contamination we observe comes from the smoothing of emission lines. To produce $\hat{\overline{C}}$ from our template $\overline{C}$, we need to emulate the same smoothing. We do this by broadening the emission lines in our template $\overline{C}$ (listed in table \ref{tab:elines}), via a parameter $\sigma_{\rm cont}$ representing the amount of the velocity dispersion in $\rm kms^{-1}$ \footnote{The velocity dispersion here is related to redshift errors.}. The width of each emission line is then broadened according to: 

\begin{equation}
    \hat{\sigma}_{\rm line} = \sqrt{\sigma_{\rm line}^2 + \left(\frac{\lambda_{\rm line} \sigma_{\rm cont}}{c}\right)^2},
\end{equation}

\noindent where $\sigma_{\rm line}$ is the intrinsic line width, $\lambda_{\rm line}$ is the rest-frame wavelength of the line, and $c$ is the speed of light in $\rm kms^{-1}$. Note that since the emission lines are normalised Gaussians, the line amplitudes are $\propto 1/\hat{\sigma}_{\rm line}$. With the broadened emission lines we can construct $\hat{\overline{C}}(\lambda, \sigma_{\rm cont})$, and obtain $\gamma(\lambda)$. As mentioned earlier, this is robust to the choice of smooth component but highly dependent on the emission line model.




\begin{figure*}
    \centering
    \includegraphics[width=\linewidth]{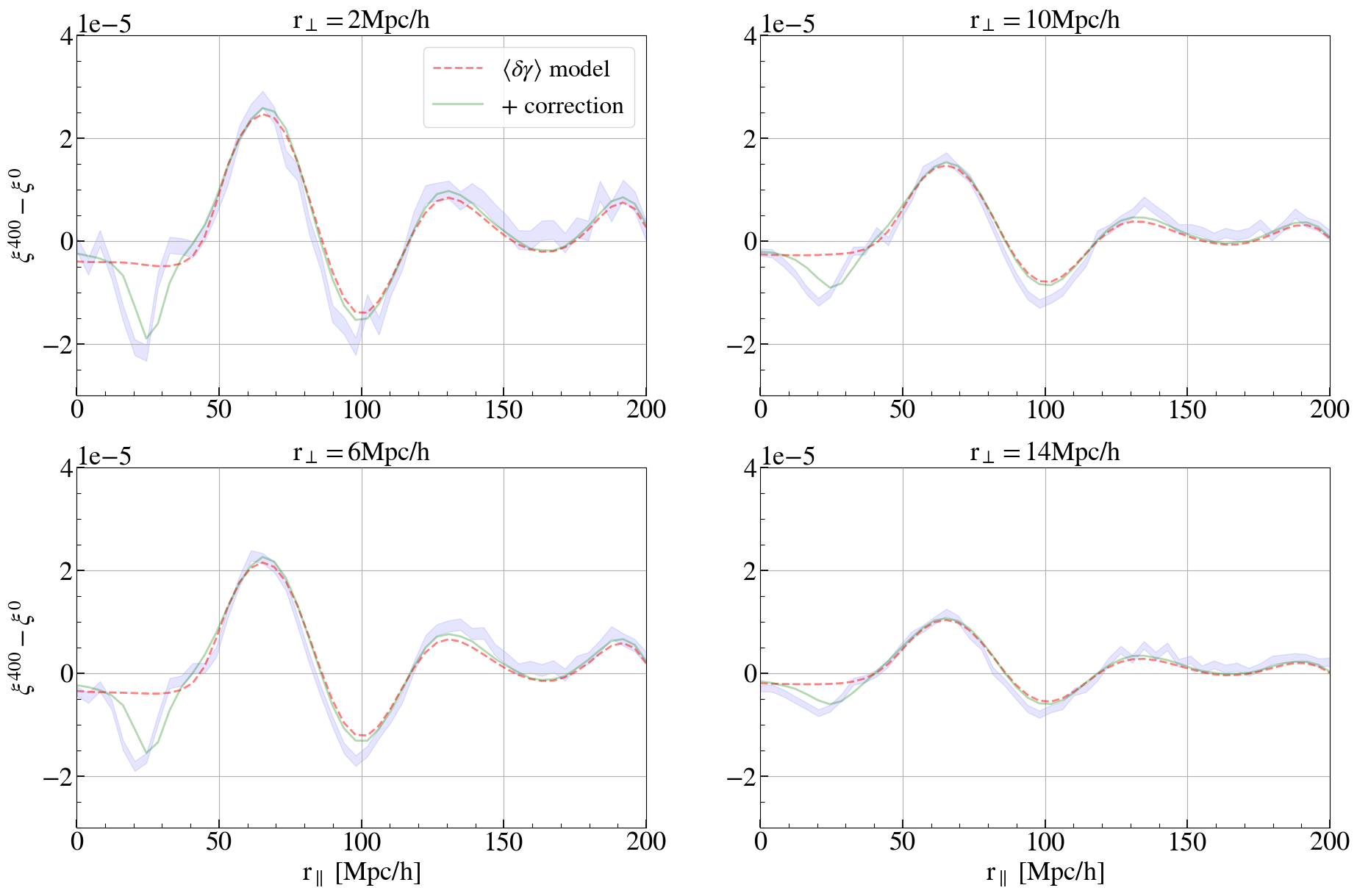}
    \caption{Direct fits to the contamination introduced by redshift errors, or the difference in the contaminated and uncontaminated \lya autocorrelation, for the first 4 bins in $r_\perp$. We plot our model with (+correction) and without ($\langle\delta\gamma\rangle$ model) the addition of small-scale correction described in appendix \ref{sec:appB}. Note we use the $\gamma(\lambda^{\rm rf})$ model described in section \ref{sec:modgamma} to compute $\langle\delta\gamma\rangle$, rather than a measurement from mocks. In this plot we show the stack of 100 DESI DR1 mocks, used throughout this analysis.}
    \label{fig:modfit}
\end{figure*}

The result of fitting our model directly to the redshift error contamination is shown in figure \ref{fig:modfit} (using a model for $\gamma$). We show fits with the small-scale correction of appendix \ref{sec:appB} (green solid line) and without (red dashed line), which we see are identical with the exception of the small-scale trough. In the full analysis (section \ref{sec:fits}) we choose not to use the small-scale correction model, to avoid biasing our results. We are still able to fit our model and recover the true input $\sigma_{\rm cont}$ without it, but may fail to capture the full impact of the trough at $\sim25\,h^{-1}$Mpc. We see that the contamination is well fit at all scales, and that it drops off quickly with transverse separation.

\subsection{Cross-correlation model}\label{sec:crosscor_mod}

In \cite{Youles} they introduced a model of continuum redshift errors for the \lya-quasar cross-correlation. The contamination in this case is characterised as: $\langle\hat{\delta}\rangle=\langle\delta\rangle-\langle \gamma(\lambda^{\rm rf})\rangle$, where $\langle\hat{\delta}\rangle=\hat{\xi}^{\rm X}$. The model of $\gamma(\lambda^{\rm rf})\rangle$ is then constructed using a similar method to the one in section \ref{sec:autocorr_mod}, resulting in: 

\begin{equation}\label{eq:xcf_cont_err_mod}
    \langle \gamma \rangle_A = A_{\rm cont}^{\rm cross} \int \int dz_{i}^{\rm Q} dz_{j}^{\rm Q} P(z_{i}^{\rm Q}) P(z_{j}^{\rm Q}) \xi^{\rm Q}(\Vec{r}^{\rm Q})\gamma_{i}, 
\end{equation}

\noindent where there is now dependence on the quasar autocorrelation $\xi^{\rm Q}$, rather than the \lya-quasar cross-correlation. To include the \cite{Youles} model in our fits, we use the same $\gamma$ model as in the autocorrelation (section \ref{sec:modgamma}). For this we have two free parameters, $\sigma_{\rm cont}$ and $A_{\rm cont}^{\rm cross}$. Note that in the cross-correlation the model of \cite{Youles} also fails to capture the behaviour of redshift errors at $\sim25\,h^{-1}$Mpc, where instead of a trough we now see a peak (visible in figure \ref{fig:mitigate_cf_xcf}). However, because we limit our cross-correlation to scales above 40$\,h^{-1}$Mpc (see section \ref{sec:fits}), the impact of this peak is much smaller. 


We ultimately combine both correlation functions in a joint fit, shown in section \ref{sec:fits}. This allows us to break degeneracies between certain nuisance parameters and key parameters like $\phi_{\rm s}$, $\alpha_{\rm s}$. As explained in \cite{CuceuAPMethod}, we can also measure the combination $f\sigma_8$ from the joint fit. We expect $\gamma$ to be the same for the auto- and cross-correlations, so we use only one parameter $\sigma_{\rm cont}$ for both. We have verified that allowing each correlation function to have a different set of parameters give perfectly consistent results.



\section{Results}\label{sec:fits}

In this section we will explain our model fitting procedure, and show the impact of redshift errors on the cosmological parameters of our full-shape analysis. In particular, we discuss how large this contamination is with respect to the precision of DESI DR1 and DR2. We will then present the full-shape fits including the redshift errors model for the auto- (section \ref{sec:autocorr_mod}) and cross-correlations (section \ref{sec:crosscor_mod}). We show results of fits to these correlations individually, and together in a joint fit which gives constraints on the combination $f\sigma_8$.

\subsection{Parameter degeneracies}
The RSD parameter of the \lya forest is given by:

\begin{equation}
    \beta_\alpha = \frac{b_{\eta,\alpha(z)} f(z)}{b_\alpha(z)} ,
\end{equation}

\noindent where $b_{\eta,\alpha}$ is the velocity divergence bias of the \lya forest and $f$ is the logarithmic growth-rate. In practice we must assume a linear matter power spectrum template \citep{Planck18}, which has a fixed normalisation proportional to $\sigma_8(z)$ - the amplitude of perturbations in 8$h^{-1}$Mpc spheres. In linear theory $f$ and $\sigma_8$ are fully degenerate \citep{Percival09}, so we are sensitive to the combinations $b_{\eta,\alpha}f\sigma_8$ and $b_\alpha \sigma_8$. Since $b_{\eta,\alpha}$ is unknown, we treat $\beta_\alpha$ as a nuisance parameter to marginalise over.

For the cross-correlation, we are sensitive to both the quasar and \lya RSD and bias parameters. Since for quasars $\beta_{\rm q} = f/b_{\rm q}$, we are sensitive to \{$b_\alpha\sigma_8,b_{\eta,\alpha}f\sigma_8,b_{\rm q}\sigma_8, f\sigma_8$\}. $f\sigma_8$ is difficult to constrain from the cross-correlation alone, since it's dependent on several biases ($b_{\eta,\alpha},b_\alpha,b_{\rm q}$). But we break these degeneracies by combining both correlation functions in a joint fit, as the autocorrelation provides precise measurements of $b_{\eta,\alpha}f\sigma_8$ and $b_\alpha \sigma_8$.

\subsection{Fits}\label{sec:fitssub}


Following the full-shape analysis of DESI year-1 \lya data \citep{LyaFS}, we restrict the autocorrelation to $r \in [25,180]\,h^{-1}$Mpc, and cross-correlation to $r \in [40,180]\,h^{-1}$Mpc. This is a conservative minimum scale cut designed to limit the impact of increasingly non-linear scales and systematics (including redshift errors). We highlight at this point that there are some key differences between the analysis we perform here, and that of \cite{LyaFS}. First of all, as mentioned in section \ref{sec:method}, \cite{LyaFS} use two additional correlation functions measured from \lya absorption in \lyb region (between 920-1020\,\AA). These are the autocorrelation of \lya absorption in the \lyb region, and its cross-correlation with quasars. We omit these correlations from our analysis for simplicity, since we would need to extend our continuum redshift errors model to incorporate this new region. We also do not expect redshift errors in the \lyb region to have a large impact overall, since these correlation functions have low statistical power relative to the two we analyse in this paper. Furthermore, in this paper we measure the AP effect separately on the broadband ($\phi_{\rm s}$) and peak component ($\phi_{\rm p}$) of the correlation function, as opposed to \cite{LyaFS} who measure the AP effect across the full correlation function with one parameter ($\phi_{\rm f}$). We opt for the former to characterise the impact of redshift errors more precisely, but it may result in a larger overall shift. Finally, we use a Gaussian prior on $L_{\rm HCD}$, compared to \cite{LyaFS} who use a wide uniform prior. We do this to make fitting our new redshift error model parameters easier, but it may also lead to slight differences in our results.

Since we are only interested in our set of cosmological parameters \{$\alpha_{\rm p},\phi_{\rm p},\phi_{\rm s},f\sigma_8$\} and the parameters of the model introduced in this paper \{$\sigma_{\rm cont}, A_{\rm cont}^{\rm auto},A_{\rm cont}^{\rm cross}$\}, we treat all other parameters as nuisance. Note that we have not included $\alpha_{\rm s}$ in our parameters of interest, since it is not simple to extract cosmological from it. 


In the autocorrelation fits, we have 15 free parameters when using the redshift errors model, and 13 without it. We fix $\beta_{\rm HCD}$ and $L_{\rm HCD}$ to values constrained in the joint fit, and set $f\sigma_8$ to value of the fiducial cosmology. For our model (equation \ref{eq:dg_final}), we also need the redshift distributions of \lya pixels and quasars in our dataset, our gamma model $\gamma(\lambda^{\rm rf})$ and an input cross-correlation function $\xi^{\rm X}$. Ideally, one would use a model (i.e. one that we fit in this analysis), but for simplicity and to reduce computing time we will use the measured cross-correlation from our mock dataset (figure \ref{fig:auto_cross_corr}). This is a reasonable approximation since the stack of mocks has very high signal-to-noise. For the DESI year-1 dataset, as we discuss in section \ref{sec:fits_to_data}, the cross-correlation is noisy and therefore is a not a good substitute for a model.

For the cross-correlation we have 16 free parameters with the redshift errors model and 14 without. We additionally fit for systematic redshift errors ($\Delta r_\parallel)$, and smoothing at large $k_\parallel$ due to redshift errors and peculiar velocities ($\sigma_{\rm v}$). However, we now fix $\sigma_\parallel$ \footnote{The line-of-sight smoothing due to limited simulation grid size.} to the joint fit value since it is degenerate with $\sigma_{\rm v}$. For this we also provide a quasar redshift distribution, the continuum smoothing function $\gamma(\lambda^{\rm rf})$ (see section \ref{sec:modgamma}) and the quasar autocorrelation $\xi^{\rm Q}$ (section \ref{sec:crosscor_mod}). Again we make a simplification by using a measurement of the quasar autocorrelation in-place of a model.

The joint fits have have 22 free parameters, combining the auto- and cross-correlations. Our full list of priors for all free parameters in the joint fit is given in table \ref{tab:priors} in appendix \ref{sec:appA}. In most cases we use a wide uniform prior, except for the HCD length-scale ($L_{\rm HCD}$) and RSD parameter ($\beta_{\rm HCD}$), which where assign Gaussian priors based on previous studies \citep{IgnasiDLAxLya,DESILYA}. Our joint fits in both cases have an effective redshift z $\sim$ 2.3. We compute our full-shape model using the \texttt{Vega}\footnote{https://github.com/andreicuceu/vega} library, which also contains a set of analysis tools. For the auto- and cross-correlation fits, we use the \texttt{Vega} interface for \texttt{Iminuit}, which performs a fast $\chi^2$ minimisation. For the joint correlations we use the nested sampler \texttt{Polychord}\footnote{https://github.com/PolyChord/PolyChordLite}\citep{Polychord} to sample the posterior space of our parameters. The latter is slower, but is able to capture any non-Gaussianity in our posterior space, and gives better error estimates than a $\chi^2$ minimisation.


\begin{figure*}
    \centering
    \includegraphics[width=1\linewidth]{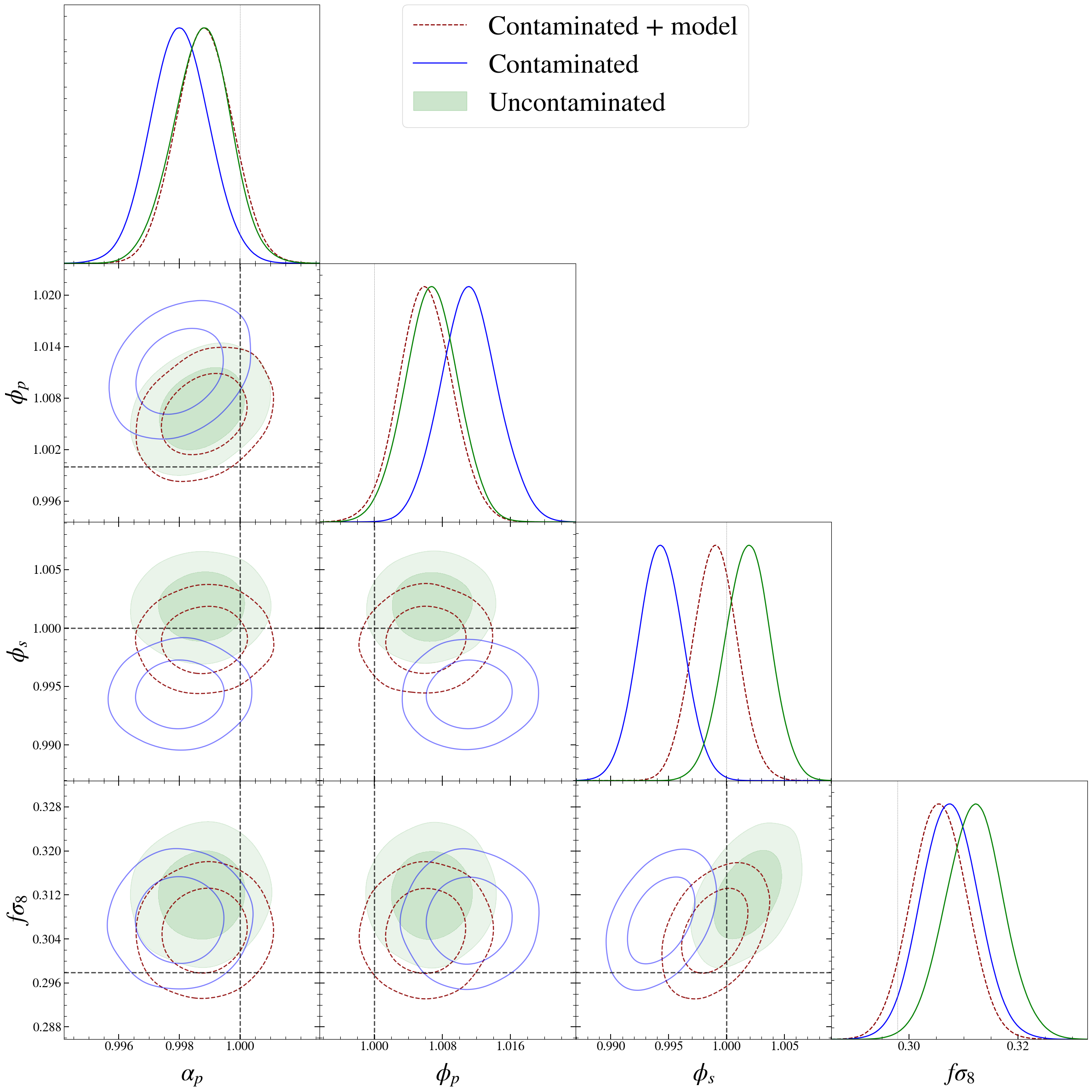}
    \caption{Full-shape posteriors from fits to a stack of 100 DESI DR1 mocks. We compare fit results from an uncontaminated (without redshift errors) dataset (green shaded), to fits on a contaminated dataset with (red dashed) and without our model (blue solid). Our fiducial cosmology is also indicated with black dashed crosshairs.}
    \label{fig:FScont}
\end{figure*}

In figure \ref{fig:FScont} we present the posteriors of $\phi,\alpha,\phi_{\rm s},f\sigma_8$, from nested sampler runs on the joint correlation function for both the uncontaminated and contaminated datasets. For the latter we include two sets of contours, for runs including (red dashed) and not including (blue solid) the redshift errors model presented in this paper. We first note that the errors on these differences are very small, because we are analysing the stack of 100 DESI year-1 mocks. Later, when looking at the full set of results in table \ref{tab:results}, we will compare our results to the precision of DESI DR1 \citep{LyaFS}. We also note that the uncontaminated contours are slightly shifted from our fiducial cosmology \citep{Planck18}, but we will not discuss this here since we are specifically concerned about the impact of redshift errors (shift from green to blue). In appendix \ref{sec:appA}, we show the full triangle plot of our joint correlation run, and discuss what effect our model has on other nuisance parameters. We see that, in general, the shifts introduced by continuum redshift errors are small compared to the precision of DESI DR1. In this paper, we consider any shift greater than $\sigma_{\rm DR1}/3$ to be significant, in-line with the requirements of \cite{DESILYA}. It should be noted however, that the shift we refer to here is only due to the impact of continuum redshift errors. In other words, this shift is not equivalent to a bias with respect to the fiducial cosmology (dashed crosshairs in figure \ref{fig:FScont}).

From figure \ref{fig:FScont}, we see $\phi_{\rm p}$ is more affected than $\alpha_{\rm p}$, but the shift is still minor. The smooth AP parameter ($\phi_{\rm s}$) is most biased by continuum redshift errors because it depends on the broadband component of our correlation functions, which is more affected by contaminants in general. The observed shift may also be smaller because of the conservative minimum separations we choose for the auto- (25\,$h^{-1}$Mpc) and cross-correlations (40\,$h^{-1}$Mpc). A small but insignificant shift occurs in the $f\sigma_8$ posterior, which is also sensitive to the broadband of the correlation functions. 

After introducing our model (red dashed contours), we see an impact on each of our posteriors. Firstly, we successfully recover the BAO peak parameters ($\alpha_{\rm p}$ and $\phi_{\rm p}$) from the uncontaminated case. This result shows that our model can be used to correct for continuum redshift errors on BAO measurements, without considering a small-scale correction (e.g. equation \ref{eq:addcorr}). 

The shift introduced by continuum redshift errors on $\phi_{\rm s}$ is reduced significantly upon introduction of our model, but still differs slightly from the uncontaminated case. As discussed in section \ref{sec:model}, our model does not capture the redshift error contamination present at $\sim25\,h^{-1}$Mpc, which is responsible for the remainder of the shift. The $f\sigma_8$ constraint is shifted very marginally further from the uncontaminated case when using our model, towards the fiducial value (black dashed crosshairs). Indeed, it seems that all of our posteriors are shifted marginally towards the fiducial value, which suggests our new model is capturing some imperfections in the existing model. However, as we will show in appendix \ref{sec:appA}, our model parameters are not correlated with the cosmological parameters of interest, thus we are fitting for some other "nuisance" effect. The shift in f$\sigma_8$ is still small and consistent with the uncontaminated case to within 1-sigma.

In table \ref{tab:results} we show the differences in the cosmological parameters of our full-shape model, between analyses on uncontaminated and contaminated mocks. The columns under "+Model" give the same differences after adding our models of the contamination (section \ref{sec:model}) to the full-shape fits. We ignore nuisance parameters constraints, except those of the model introduced in this paper, which we show in table \ref{tab:modelresults}. As noted earlier, we are analysing the stack of 100 DESI DR1 mocks with high signal-to-noise. For that reason, we include the projected precision of the DESI year-1 full-shape analyses on data in the final column of the table (see \cite{LyaFS}).

The shift on $\alpha_{\rm p}$ (isotropic BAO parameter, equation \ref{eq:BAOiso}) is never more than $\sim10\%$ of the DR1 68\% confidence level ($0.1\sigma_{\rm DR1}$). For $\phi_{\rm p}$ (AP effect on the BAO peak, section \ref{sec:BAOAP}) and the growth-rate, we also see shifts of $\sim0.1\sigma_{\rm DR1}$ in the joint fit. For each of the latter 3 parameters, the shifts introduced by redshift errors are notably consistent with zero to $\sim1\sigma$. However, the shift on $\phi_{\rm s}$ is $0.44(\pm0.13)\sigma_{\rm DR1}$, which is larger than the $\sigma_{\rm DR1}/3$ criterion we set earlier. Due to large computing time, errors on these shifts are estimated from the stack of mocks, rather than the distribution of shifts across the 100 individual mocks. This approximation is justified because the shift is systematic in nature, and we are using the same mock realisations on both sides of the comparison, so the impact of noise is minimised.

While we have only included DESI DR1 projections, it is important to note that the relative sizes of the shifts shown here will increase significantly for the DESI DR2 and future datasets. The DR2 constraints for example should be approximately $\sim$40\% tighter than DR1, making the $\phi_{\rm s}$ shift $\sim0.7\sigma_{\rm DR2}$. It is also important to stress that this shift is not a bias with respect to the fiducial cosmology, rather it characterises the impact of a single contaminant. It may be the case that parametrising the AP effect over the full correlation function instead of the peak and broadband components separately, as is done in \cite{LyaFS}, reduces the impact of redshift errors. 

The next 3 columns (+ model) show the same differences as the first 3, but including the continuum redshift errors model. For the peak component parameters ($\alpha_{\rm p},\phi_{\rm p}$), our model is very successful in removing the bias for all 3 fit cases. For $\phi_{\rm s}$, our model successfully reduces the bias in the joint case to $0.19(\pm0.13)\sigma_{\rm DR1}$, now consistent with zero to 1.5$\sigma$. Our model has almost no impact on $f\sigma_8$, but the shift is small $\sim0.1\sigma_{\rm DR1}$ and consistent with zero to $\sim$1$\sigma$. The parameter values of our model constrained in the analysis are shown in the section below. 

As previously mentioned, we have a single set of parameters controlling the shape of $\gamma$: $\sigma_{\rm cont}$, $A^{\rm auto}_{\rm cont}$ (autocorrelation amplitude) and $A^{\rm cross}_{\rm cont}$ (cross-correlation amplitude), which have values given in table \ref{tab:modelresults}. In each case we are able to recover the input redshift error (400\,$\rm kms^{-1}$) to within a 1$\sigma$ confidence interval, and have strong detections of the other 3 parameters. The cross-correlation amplitude parameter is consistent between the cross and joint cases; however, this is not true for the auto-correlation amplitude. The tension between the auto and joint cases may reflect the incompleteness of the model at small-scales. For each of these parameters we apply wide uniform priors, shown in table \ref{tab:priors} with the rest of the parameters in the joint fit. 

The $\chi^2$ probabilities of the analyses both with and without our model are very low. This is expected, since we are fitting an extremely high signal-to-noise dataset with a linear theory model (section \ref{sec:corrmod}) that is suboptimal. Furthermore, it has been shown in \cite{LyaFS} that our model is suitable for at least the level of precision of DESI DR1. We see a large improvement in $\chi^2$ value when using our model in the autocorrelation ($\Delta\chi^2=655$, 1564 bins, 15 parameters), the cross-correlation ($\Delta\chi^2=726$, 3030 bins, 16 parameters) and the joint ($\Delta\chi^2=970$, 4594 bins, 22 parameters) correlation runs.


\begin{table*}
  \begin{tabular}{|c|ccc|ccc|c}
    \hline
    \multirow{2}{*}{Parameter} &
      \multicolumn{3}{c|}{Contaminated fit} &
      \multicolumn{3}{c|}{+ Model} &
      \multirow{1}{*}{$\sigma_{\rm Y1}$} \\
    
     & auto & cross & joint & auto & cross & joint & \\
    \hline
    $10^3\Delta\alpha_{\rm p}$ & 0.1 $\pm$ 1.4 & 1 $\pm$ 1.2 & 0.8 $\pm$ 1.0 & -0.2 $\pm$ 1.4 & 0.03 $\pm$ 1.2 & -0.1 $\pm$ 0.9  & 11 \\
    $10^3\Delta\phi_{\rm p}$  & -8 $\pm$ 5 & -4 $\pm$ 4  & -4 $\pm$ 3 & -5 $\pm$ 5 & 1 $\pm$ 4  & -0.7 $\pm$ 3  & 38 \\
    $10^3\Delta\phi_{\rm s}$  & 4 $\pm$ 2.4 & 10 $\pm$ 4 & -7 $\pm$ 2 & 3 $\pm$ 2.4 & 10 $\pm$ 3 & -3 $\pm$ 2  & 16 \\
    $10^2\Delta{f}\sigma_8$ & - & - & 0.5 $\pm$ 0.5 & - & - & 0.6 $\pm$ 0.5 & 6.2 \\
    \hline
  \end{tabular}
  \caption{Differences in parameter constraints on fits to mocks with and without redshift errors. We fit a stack of 100 DESI DR1 contaminated mocks, without (Contaminated fit) and with (+ Model) our continuum redshift errors model, and measure the shift with respect to the uncontaminated set. The last column shows the projected errors (68\% confidence interval) on the same parameters for the joint constraints from DESI DR1 dataset. Note that the values presented here are shifts introduced solely by continuum redshift errors, not the bias with respect to the fiducial cosmology.}
  \label{tab:results}
\end{table*}

    

\begin{table*}
  \begin{tabular}{c|ccc}
    \hline
    {Parameter} & auto & cross & joint  \\
    \hline
    $\sigma_{\rm cont} [\rm kms^{-1}]$  & 401 $\pm$ 19  & 406 $\pm$ 11 & 411 $\pm$ 18 \\
    $A_{\gamma}^{\rm auto}$ & -6.6 $\pm$ 0.4 & - & -8.9 $\pm$ 0.4  \\
    $A_{\gamma}^{\rm cross}$ & -  & -6.2 $\pm$ 0.3 & -5.8 $\pm$ 0.2 \\
    \hline
  \end{tabular}
  \caption{Constraints on free parameters of the model we introduce to capture continuum redshift errors (section \ref{sec:model}), from a fit to a stack of 100 DESI DR1 contaminated mocks. We show the constraints (section \ref{sec:model}) for the auto-, cross- and joint correlations separately. The input amount of redshift error was $\sigma=400 \, \rm kms^{-1}$, which we recover to within 1$\sigma$ in the each case.}
  \label{tab:modelresults}
\end{table*}

\section{Discussion}\label{sec:fits_to_data}

In the previous section, we showed and discussed the results of fits to synthetic data sets with the model introduced in this paper. In section \ref{sec:cont_real_data} we will discuss the application of this model to real data, and highlight potential differences with the synthetic datasets we use. We will also show in section \ref{sec:mitigate_cont}, that the effect of continuum redshift errors can be mitigated by removing pairs where the quasar autocorrelation (for contamination in the \lya-quasar cross-correlation) or the \lya-quasar cross-correlation (for contamination in the \lya autocorrelation) is large.


\subsection{Contamination on real data}\label{sec:cont_real_data}



In \cite{Youles}, they found a strong dependence of $\langle \gamma \rangle$ on small-scale quasar clustering. Likewise, we verify that $\langle \delta \gamma \rangle$ has strong dependence on the small-scale cross-correlation function. This is discussed further in section \ref{sec:mitigate_cont}. We also find that the quasar clustering in our lognormal mocks deviates more from linear theory at small-scales than more realistic n-body simulations (see figure 6 of \cite{Youles}), which is likely to increase the amount of contamination relative to real data. 

As mentioned in section \ref{sec:fitssub}, we use a measurement of the cross-correlation in our $\langle \delta \gamma \rangle$ model, and a measurement of the quasar autocorrelation in $\langle \gamma \rangle$. For our mock datasets, where we have high signal-to-noise these measurements are smooth, and are a better alternative to using a model of the respective correlation functions. However, in e.g. DESI DR1, our correlations are relatively noisy and, in the case of $\hat{\xi}^X$, and are themselves affected by redshift errors. It is also important to note that because of our strong dependence on small-scale clustering, using a linear model of $\xi^{\rm X}$ in equation \ref{eq:dg_final} (or $\xi^{\rm Q}$ in equation \ref{eq:xcf_cont_err_mod}) would likely be a bad approximation. For these reasons, we choose to leave a full analysis on data for future work where we have a higher signal-to-noise data set, or a cross-correlation (and quasar autocorrelation) model that is realistic at small-scales. At the level of precision of DESI DR1, our model is also not yet necessary to perform an un-biased analysis (i.e. \cite{LyaFS}).

Another contributing factor in our analysis realism is our synthetic spectra, or specifically the quasar continua that we use. As we discussed in section \ref{sec:method}, we generate quasar spectral templates using convolutions of power law smooth components, and emission lines (table \ref{tab:elines}) from the composite spectrum of BOSS quasars \citep{SIMQSO,BOSSComposite}. We sample from a Gaussian distribution of power law slopes, tuned to have mean and scatter that better reflected the eBOSS DR16 dataset \citep{dmdb2020}. Likewise, for the \lya forest emission lines we sample from a Gaussian distribution of EWs, tuned on BOSS DR9 data. The distribution of emission line EWs is particularly important for our study, since the dominant contribution to the contamination is around the positions of these. If, for example, the mean EW in our spectra was higher than in e.g. DESI DR1, we would observe a higher level of contamination. Also, since the scales of the spurious correlation features are directly related to the rest-frame wavelength of the emission lines in our \lya forests (see section \ref{sec:evolwav}), we are sensitive to the relative strength of individual lines. 


The way we introduce redshift errors into our spectra should also reflect the magnitude and nature of redshift errors in real data. However, this is difficult to emulate because there are several ways in which redshift errors are produced. We add errors drawn from a Gaussian with dispersion $\sigma=400 \, \rm kms^{-1}$ to each quasar in our sample, representing errors which might occur in our redshift estimation pipeline \texttt{Redrock}. However, as mentioned in section \ref{sec:intro}, it is also typical for emission lines to shift away from the systemic redshift of the quasar. The level of shift depends on the emission line, but is larger in general for broad, high-ionisation lines like CIV, SiIV and CIII. Therefore, the level of error depends on the set of lines being used to estimate the quasar redshift. There has been progress made in this area recently, with work on improving spectral templates in DESI \citep{Bault24,AllysonTemplates}.

There is further complication due to the fact that the contaminating lines within the \lya forest are high-ionisation. If these are strongly correlated with the high-ionisation lines used to estimate quasar redshift, then there would be little impact on the mean continuum. The latter point requires further studies like the one of \cite{Shen16,Bault24,AllysonTemplates}, but is difficult in practice because \lya forest lines are less prominent, and visible at a higher redshift ($z \gtrsim 2.1$). One could improve the way we add redshift errors to mocks using these studies to input typical relative line shifts into our spectra, and use a redshift estimator (i.e. \texttt{Redrock}) to estimate our redshifts.

\subsection{Mitigating contamination in real data}\label{sec:mitigate_cont}

\cite{Youles} identified that the level of redshift error contamination in the cross-correlation between \lya and quasars, depends on the correlation between the host quasar of the forest and the correlating quasar. This correlation is larger for neighbouring pairs of quasars, and avoiding these configurations significantly reduces the contamination. Here we extend this further by explicitly removing \lya-quasar pairs where the forest host-quasar and correlating quasar are separated by less than a given separation. The situation is similar for the \lya autocorrelation, except that, as we showed in section \ref{sec:autocorr_mod}, the contamination is now dependent on the \lya-quasar cross-correlation. 

\begin{figure}
    \centering
    \includegraphics[width=\linewidth]{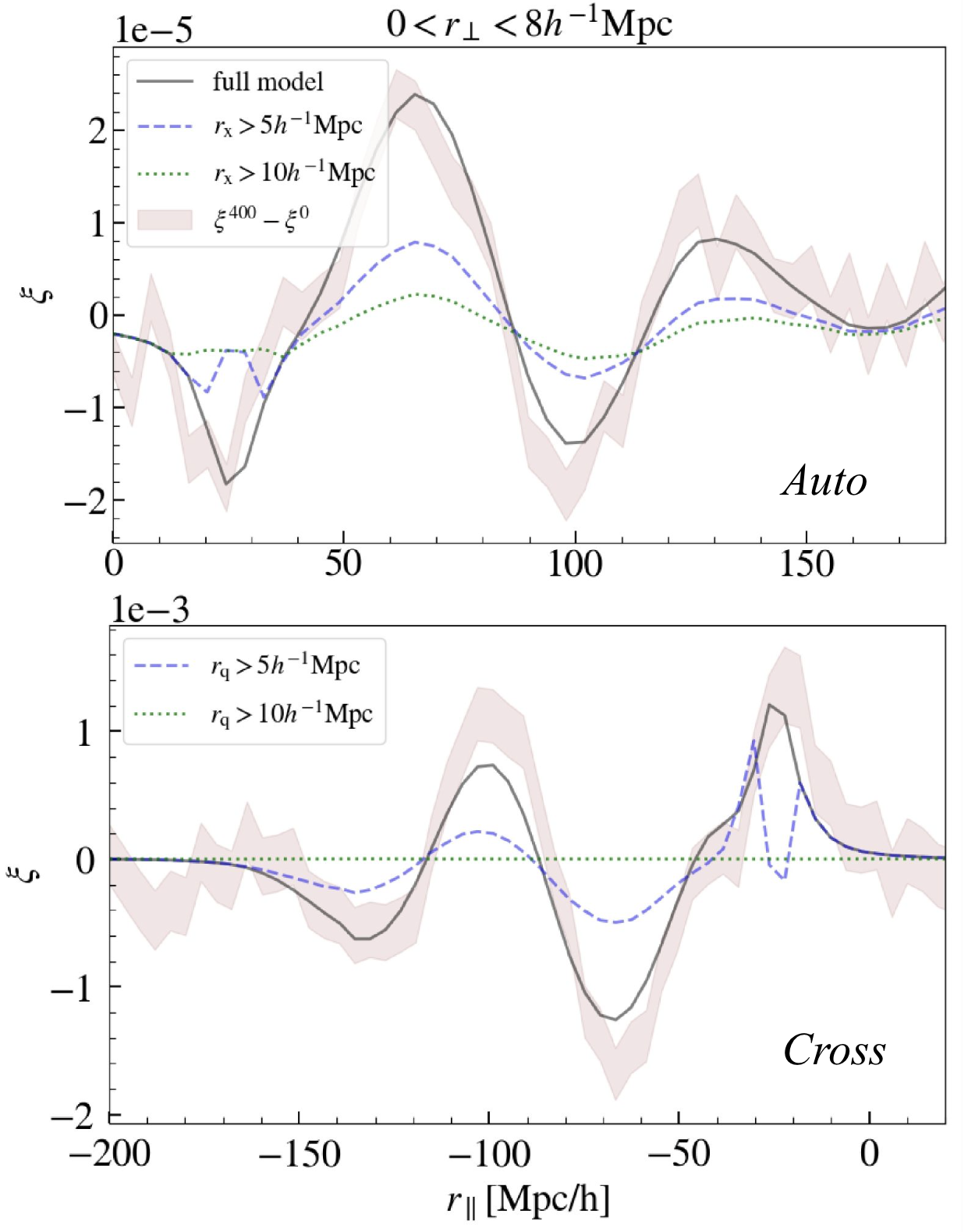}
    \caption{(Top) the contamination in the \lya autocorrelation introduced by redshift continuum errors and our model of it (black solid). Also plotted is the same model evaluated only for separations $r_{\rm X}$ greater than 5\,$h^{-1}$Mpc (blue dashed) and 10\,$h^{-1}$Mpc (green dotted), where $r_{\rm X}$ is the separation between one pixel and the host quasar of the other pixel. (Bottom) continuum redshift errors contamination in the \lya-quasar cross-correlation, over-plotted with the model of the same contamination. The model is evaluated with the same cuts as the autocorrelation, but with $r_{\rm q}$ now referring to the separation between a correlating quasar and the host quasar of the correlating \lya forest pixel. In both cases we have used the small-scale correction of appendix \ref{sec:appB}.}
    \label{fig:mitigate_cf_xcf}
\end{figure}

The result of removing contributions to our model from pairs where the host quasar of a \lya pixel and the correlating quasar are separated by less than $r_{\rm q}$, is shown in figure \ref{fig:mitigate_cf_xcf} (bottom). We choose two cuts at 5\,$h^{-1}$Mpc and 10\,$h^{-1}$Mpc, which removes only $\sim0.1-0.3\%$ of our total number of pairs, but $\sim10\%$ in the first 5 transverse bins (up to $r_\perp\sim20$ $h^{-1}$Mpc). From figure \ref{fig:mitigate_cf_xcf}, we see that by 10\,$h^{-1}$Mpc, the contamination is effectively negligible.

In the top panel of figure \ref{fig:mitigate_cf_xcf} we show an analogous test with the contamination in the \lya autocorrelation. Here we remove pairs where the host quasar of one forest is separated from the correlating pixel in the other forest by less than $r_{\rm X}=5,10\,h^{-1}$Mpc. In this case, we see a similar same level of success in removing the contamination. The fraction of correlating pairs that is removed in this test is also sub-percent level overall, but $\sim10\%$ in the most line-of-sight bins. 

These tests were undertaken in \cite{Casas25,LYADR2} and shown to have very little impact on the precision of BAO constraints, where the most line-of-sight bins are at much smaller separation than the BAO feature. However, there was a significant impact on the \lya bias, which may suggest this approach is not suitable for full-shape analyses. We leave this to be studied in future work, as a possible alternative to the model we present in this paper.




\section{Summary}\label{sec:summary}
Random errors in our quasar redshift measurements smooth the mean continuum used to estimate the \lya flux transmission field. This smoothing gives rise to spurious features across the length of the \lya forest, which in-turn distort the full-shape of the \lya autocorrelation and its cross-correlation with quasars. 

In this paper we presented a model of these spurious correlations for the \lya autocorrelation function, building upon work in \cite{Youles}, where they presented an equivalent model for the \lya-quasar cross-correlation. We then created a model for the distortion as a function of rest-frame wavelength ($\gamma(\lambda^{\rm rf},\sigma_{\rm cont})$), and introduced three new parameters to our standard full-shape model. These control the amount of smoothing introduced by redshift errors ($\sigma_{\rm cont}$), and the amplitude of the effect in the auto- ($A_{\gamma}^{\rm auto}$) and cross-correlation functions ($A_{\gamma}^{\rm cross}$).

We show using a high signal-to-noise synthetic dataset that continuum redshift errors shift the measurements of the growth rate ($f\sigma_8$), isotropic BAO parameter ($\alpha_{\rm p}$), anisotropic BAO parameter ($\phi_{\rm p}$) and anisotropic broadband parameter ($\phi_{\rm s}$) to varying degrees, with respect to a dataset which does not contain redshift errors. At the level of precision of DESI DR1, this shift is relatively minor in the first 3 parameters ($\sim$10\%). However, we find that $\phi_{\rm s}$ is shifted from the uncontaminated dataset (without redshift errors) value by 0.44$(\pm0.13)\sigma_{\rm DR1}$, where $\sigma_{\rm DR1}$ is projected the 68\% confidence region of the DESI DR1 constraint. Note that this relative shift will increase as analyses on future datasets (e.g. DESI DR2) improve the constraining power on these parameters. We demonstrated that our model reduces the shift introduced to $\phi_{\rm s}$ by $\sim 60\%$, and completely removes any shift on $\alpha_{\rm p}$ and $\phi_{\rm p}$. We also recover to within 1$\sigma$ the input dispersion of the Gaussian distribution of redshift errors we add (400\,kms$^{-1}$), and strongly detect each of the amplitude parameters of the model. 

On real data, one should use models of the \lya-quasar cross-correlation and the quasar autocorrelation as inputs for the redshift errors model in the autocorrelation ($\langle \delta \gamma \rangle$) and the cross-correlation ($\langle \gamma \rangle$) respectively. This is because measurements of these functions are noisy compared to the high signal-to-noise stack of mocks we used throughout this work. We also show that continuum redshift error contamination could be mitigated in real data by removing pairs in the cross-correlation where the quasar autocorrelation is strongest, and removing pairs in the autocorrelation where the cross-correlation is strongest. In section \ref{sec:mitigate_cont}, we showed that removing pairs where $r_{\rm q}, r_{\rm X}$\footnote{The separation between quasars in the quasar auto-correlation, and \lya pixels and quasars in the cross-correlation respectively.}$< 10\,h^{-1}$Mpc is enough to make $\langle \gamma \rangle$ and $\langle \delta \gamma \rangle$ negligible. This corresponds to $\sim0.3\%$ of total correlating pairs, but $\sim10\%$ of the most line-of-sight (up to $r_\perp\sim20\,h^{-1}$Mpc) pairs, meaning this cut may be suitable for removing the effect of redshift errors on BAO constraints, but not for full-shape analyses.

\section*{Acknowledgements}
CG is partially supported by the Spanish Ministry of Science and Innovation (MICINN) under grants PGC-2018-094773-B-C31 and SEV-2016-0588. AC acknowledges support provided by NASA through the NASA Hubble Fellowship grant HST-HF2-51526.001-A awarded by the Space Telescope Science Institute, which is operated by the Association of Universities for Research in Astronomy, Incorporated, under NASA contract NAS5-26555. AFR from the European Union’s Horizon Europe research and innovation programme (COSMO-LYA, grant agreement 101044612). AFR acknowledges financial support from the Spanish Ministry of Science and Innovation under the Ramon y Cajal program (RYC-2018-025210) and the PGC2021-123012NB-C41 project. IFAE is partially funded by the CERCA program of the Generalitat de Catalunya.

This material is based upon work supported by the U.S. Department of Energy (DOE), Office of Science, Office of High-Energy Physics, under Contract No. DE–AC02–05CH11231, and by the National Energy Research Scientific Computing Center, a DOE Office of Science User Facility under the same contract. Additional support for DESI was provided by the U.S. National Science Foundation (NSF), Division of Astronomical Sciences under Contract No. AST-0950945 to the NSF’s National Optical-Infrared Astronomy Research Laboratory; the Science and Technology Facilities Council of the United Kingdom; the Gordon and Betty Moore Foundation; the Heising-Simons Foundation; the French Alternative Energies and Atomic Energy Commission (CEA); the National Council of Humanities, Science and Technology of Mexico (CONAHCYT); the Ministry of Science, Innovation and Universities of Spain (MICIU/AEI/10.13039/501100011033), and by the DESI Member Institutions: \url{https://www.desi.lbl.gov/collaborating-institutions}. Any opinions, findings, and conclusions or recommendations expressed in this material are those of the author(s) and do not necessarily reflect the views of the U. S. National Science Foundation, the U. S. Department of Energy, or any of the listed funding agencies.

The authors are honored to be permitted to conduct scientific research on I'oligam Du'ag (Kitt Peak), a mountain with particular significance to the Tohono O’odham Nation.

\section*{Data Availability}
The data points from the figures in this paper will be available in a Zenodo repository when it is accepted for publication.



\bibliographystyle{mnras}
\bibliography{main}


\appendix
\section{Full model results}\label{sec:appA}
In section \ref{sec:fits} we showed the posteriors of our set of cosmological parameters from nested sampler runs, for three different cases: uncontaminated (no redshift continuum errors, baseline model), contaminated (baseline model) and contaminated (baseline + redshift errors model). We showed that the model introduced in this paper (see section \ref{sec:model}) successfully removed the effect of continuum redshift errors on the BAO peak parameters ($\phi_{\rm p}$,$\alpha_{\rm p}$). It also removes the majority of the shift on $\phi_{\rm s}$, but has little impact on $f\sigma_8$. 

In table \ref{tab:priors}, we show the priors on parameters in the joint fit. In the first section, we show the cosmological parameters of interest, and in the second we show the nuisance parameters of the baseline model. In the final section, we show the parameters introduced in this paper to capture the effect of continuum redshift errors. For most of the parameters, we choose conservative uniform priors, except in the case of $\beta_{\rm HCD}$ and $L_{\rm HCD}$, where we follow \cite{DESILYA} and place more informative priors based on HCD studies to improve the performance of the sampler. The priors placed on parameters in our model are physically motivated, i.e. we can't have negative velocity dispersions, or positive values for the amplitude of the effect (see section \ref{sec:model}). The value of $\sigma_{\rm cont}$ directly reflects the size of the redshift errors in our dataset, meaning the upper bound of 2000\,$\rm kms^{-1}$ would be very extreme. 

\begin{table}
    \centering
    \begin{tabular}{|c|c|c|c|c|c|c}
       {Parameter} & {Prior}\\
     \hline
     \hline
    $\alpha,\phi$* & $\mathcal{U}$[0.01,2]\\  
    $f\sigma_8$ & $\mathcal{U}$[0,2]\\
    \hline
    
    $b_{\alpha}$ & $\mathcal{U}$[-2, 0]\\
    $\beta_{\alpha}$ & $\mathcal{U}$[0, 5]\\
    $b_{\rm q}$ & $\mathcal{U}$[0, 10]\\
    $\sigma_{\rm v}$ & $\mathcal{U}$[0, 15]\\
    $\Delta r_\parallel$ & $\mathcal{U}$[-3, 3]\\
    $b_{\rm HCD}$ & $\mathcal{U}$[-0.2, 0]\\
    $\beta_{\rm HCD}$ & $\mathcal{N}$[0.5, 0.09]\\
    $L_{\rm HCD}$ & $\mathcal{N}$[5, 1]\\
    $b_{\rm SiII(1190)}$ & $\mathcal{U}$[-0.02, 0.02]\\
    $b_{\rm SiII(1193)}$ & $\mathcal{U}$[-0.02, 0.02]\\
    $b_{\rm SiIII(1207)}$ & $\mathcal{U}$[-0.02, 0.02]\\
    $b_{\rm SiII(1260)}$ & $\mathcal{U}$[-0.02, 0.02]\\
    $\sigma_\parallel$ & $\mathcal{U}$[0, 10]\\
    $\sigma_\perp$ & $\mathcal{U}$[0, 10]\\
    \hline
    $\sigma_{\rm cont}$ & $\mathcal{U}$[0, 2e3]\\
    $A_{\rm cont}^{\rm auto}$ & $\mathcal{U}$[-50,0]\\
    $A_{\rm cont}^{\rm cross}$ & $\mathcal{U}$[-50,0]\\

    \end{tabular}
    \caption{Priors on free parameters used in the joint fits to our stack of 100 DESI DR1 contaminated mocks. *Recall that there are two $\alpha$ and two $\phi$ parameters, corresponding to the BAO peak and broadband of the correlation function.}
    \label{tab:priors}
\end{table}

\begin{figure*}
    \centering
    \includegraphics[width=1\linewidth]{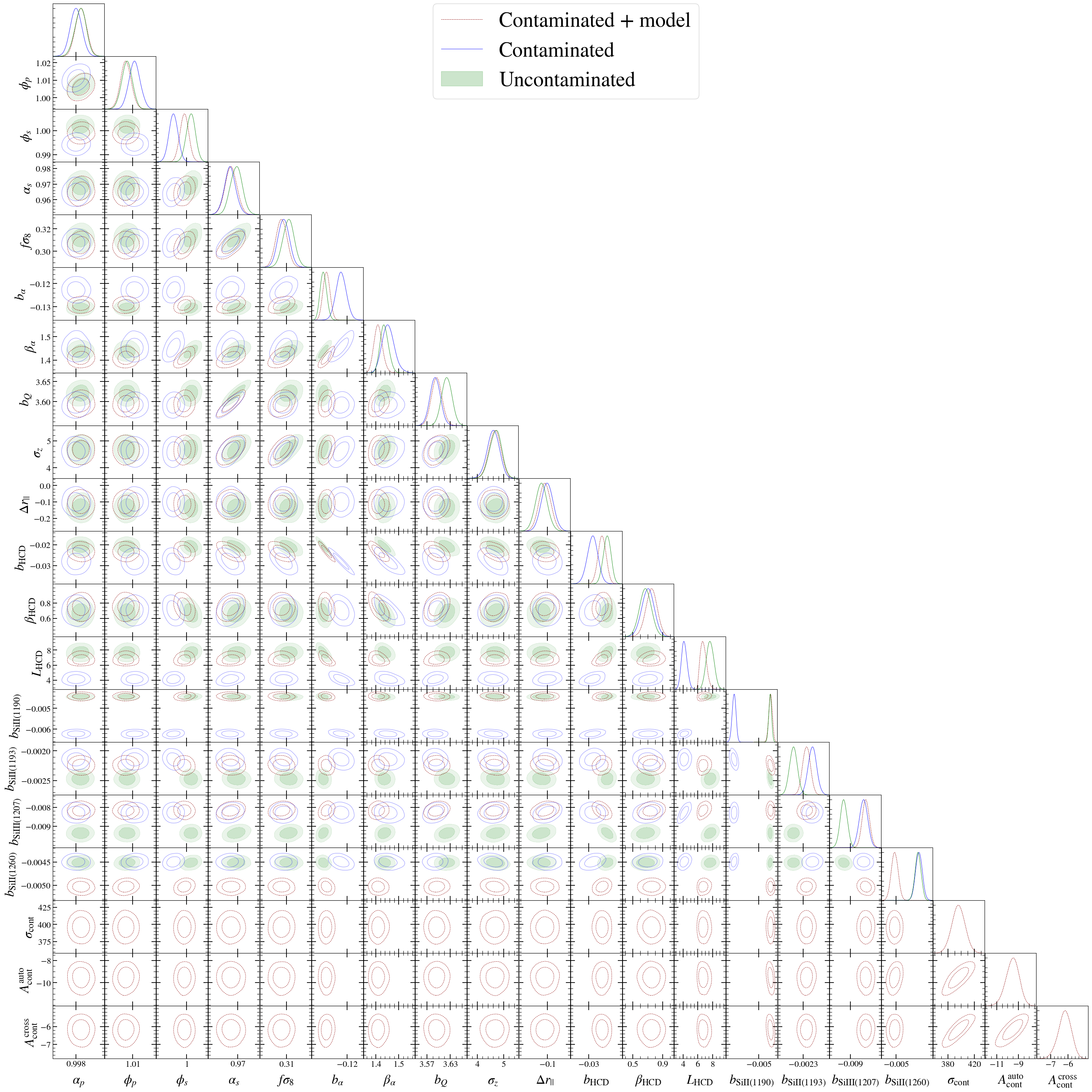}
    \caption{Posteriors of all free parameters in our sampler runs on the joint auto+cross correlation function, measured from a stack of 100 DESI DR1 mocks. The uncontaminated (without continuum redshift errors) run is shown in green, and the contaminated runs with and without the redshift errors model are shown in dark red and blue respectively. We exclude the smoothing parameters $\sigma_\parallel$ and $\sigma_\perp$, since they marginalise the effect of grid size in our simulated data sets, and are not relevant to real data.}
    \label{fig:fullcont}
\end{figure*}

In figure \ref{fig:fullcont}, we show the triangle plot of posteriors for each parameter in our sample runs. Looking first at the parameters introduced in this paper ($A^{\rm cross}_{\rm cont},A^{\rm auto}_{\rm cont},\sigma_{\rm cont}$), we see that they are mostly uncorrelated with all of the other parameters. There is a correlation between $\sigma_{\rm cont}$ and the amplitude parameters ($A^{\rm cross}_{\rm cont}, \, A^{\rm auto}_{\rm cont}$), due to the fact that broadening the (Gaussian) emission line profiles in the \lya forest also reduces their amplitude. We can also see that while our model successfully reduces the bias on the key cosmological parameters, it often fails to recover the value of other nuisance parameters in the uncontaminated case, and in one case ($b_{\rm SiII(1206)}$), is significantly more discrepant than the fit to the contaminated dataset without our model. One of the main reasons for both of these issues is the large parameter space with, in some cases, high levels of degeneracy. These differences in nuisance parameters result in the very marginal over-fitting we see in figure \ref{fig:FScont}, but really only serve to indicate that there are some systematic signals that we do not account for with our previous baseline. 
\section{Mean continuum expansion}\label{sec:appB}

For a forest of a given quasar q, the delta field is evaluated at $\lambda$ in the observed frame. The mean continuum $\overline{C}$ is a function of rest-frame, which we can write as a function of $\lambda$ and $z_{\rm q}$ as:

\begin{align}
    1 + \delta(\lambda) = \frac{f(\lambda)}{\hat{\overline{C}}(\frac{\lambda}{1+\hat{z}_{\rm q}})(a_{\rm q} + b_{\rm q}\Lambda)} ,
\end{align}

\noindent where $\Lambda$ is a function of $\log\lambda$ given in equation \ref{eq:meancontdelta}, and $\hat{\overline{C}}$ is the measured mean continuum in the presence of redshift errors. We can see from this equation that $\delta$ will be contaminated by the smoothed $\hat{\overline{C}}$, but also that the conversion from $\lambda^{\rm rf}$ to $\lambda$ will be incorrect, given we measure a quasar redshift that has some error, $\hat{z}_{\rm q} = z_{\rm q} + \epsilon$. We can therefore expand the mean continuum about the true redshift as:

\begin{align}
    \label{eq:expand_C}
    \hat{\overline{C}}(\hat{\lambda}^{\rm rf}) = \hat{\overline{C}}({\lambda}^{\rm rf}) + \epsilon\frac{d\hat{\overline{C}}}{dz_{\rm q}} + \frac{\epsilon^2}{2}\frac{d^2\hat{\overline{C}}}{dz_{\rm q}^2} + \hspace{+0.1cm} ...
\end{align}

\noindent Now, we re-write the differential with redshift as:

\begin{align}
    \label{eq:diff_Cz}
    \frac{d\hat{\overline{C}}}{dz_{\rm q}} = \frac{d\lambda^{\rm rf}}{dz_{\rm q}}\frac{d\hat{\overline{C}}}{d\lambda^{\rm rf}} = -\frac{\lambda^{\rm rf}}{1+z_{\rm q}}\frac{d\hat{\overline{C}}}{d\lambda^{\rm rf}}.
\end{align}

\noindent Substituting equations \ref{eq:expand_C} and \ref{eq:diff_Cz} into equation \ref{eq:gamma}, we can derive a new expression for our measured transmission field $\hat{\delta}_{\rm q}$:

\begin{align}
    \hat{\delta}_{\rm q} = \delta_{\rm q} - \gamma + \gamma_{\rm z} ,
\end{align}

\noindent where as usual we ignore 2nd order terms, and

\begin{align}\label{eq:gammaz}
    \gamma_{\rm z} = \frac{\lambda^{\rm rf}\epsilon}{\overline{C}(1+z_{\rm q})}\frac{d\hat{\overline{C}}}{d\lambda^{\rm rf}}.
\end{align}

\noindent Now, as we did in section \ref{sec:model}, we propagate our expression for our measured delta through to the cross- and autocorrelation functions:

\begin{align}
    \label{eq:cf_expanded}
     \langle \hat{\delta} \rangle & = \langle \delta \rangle - \langle \gamma \rangle + \langle \gamma_{\rm z}\rangle\\ 
    \langle \hat{\delta}\hat{\delta}  \rangle & = \langle \delta\delta \rangle - 2\langle \delta\gamma \rangle +  2\langle \delta\gamma_{\rm z}\rangle         - 2\langle\gamma\gamma_{\rm z}\rangle + \langle\gamma\gamma\rangle + \langle\gamma_{\rm z}\gamma_{\rm z}\rangle.
\end{align}

\begin{figure}
    \centering
    \includegraphics[width=1\linewidth]{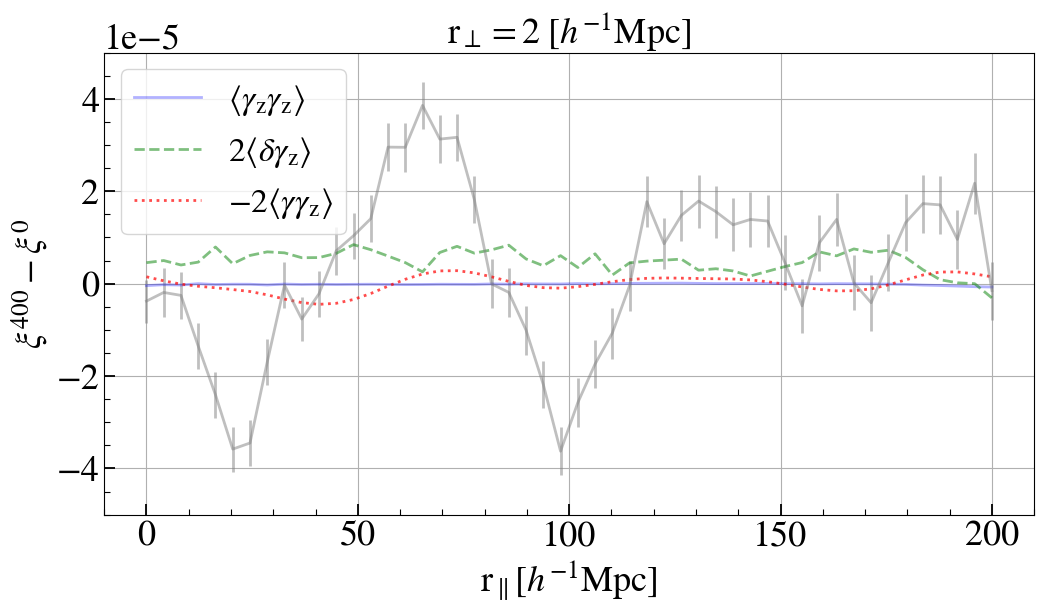}
    \caption{Contaminating terms in the \lya autocorrelation, due to rest-frame wavelength grids shifts in the presence of quasar redshift errors. $\gamma_{\rm z}$ is defined in equation \ref{eq:gammaz}, and is the result of expanding the measured mean continuum $\hat{\overline{C}}$ about small redshift errors $\epsilon$. The contamination in the first transverse bin is shown in grey, as a function of line-of-sight separation ($r_\parallel$).}
    \label{fig:zevol_terms}
\end{figure}

\noindent Modelling the additional terms analytically may prove to be difficult, but as we did for figure \ref{fig:dg_measured}, we can use mock datasets to measure each term in equation \ref{eq:cf_expanded}. In figure \ref{fig:zevol_terms}, From this measurement, we see that only $\langle\delta\gamma_{\rm z}\rangle$ is non-negligible, and that this is roughly constant across $r_\parallel$ for the usual range of $r_\perp (\lesssim 20\,h^{-1}$Mpc). An explanation for the trough feature $\sim25\,h^{-1}$Mpc is therefore still required, where perhaps higher-order terms cannot be ignored.


\subsection{Small-scale contamination model}

Here, we try to model the 25\,$h^{-1}$Mpc contaminating trough feature of our fiducial analysis (see e.g. figure \ref{fig:dg_measured}) empirically. We do this by adding an extra term to the original $\gamma$ function (equation \ref{eq:gamma}):

\begin{equation}\label{eq:addcorr}
    \gamma(\lambda^{\rm rf}) = \overline{\gamma_0}(\lambda^{\rm rf},\sigma_{\rm cont}) + \frac{a_1}{(\lambda^{\rm rf} - \lambda^{\rm rf}_{\rm 1})^2} ,
\end{equation}

\noindent where the function is offset by $\lambda^{\rm rf}_{\rm 1}=1207.6$\,\AA, and $\overline{\gamma_0}$ is the function we model in the previous section. This value was constrained in a direct fit to the difference in the \lya autocorrelation with and without continuum redshift errors, together with $a_1$, the redshift error parameter ($\sigma_{\rm cont}$, section \ref{sec:autocorr_mod}) and a free amplitude. Note this function diverges for $\lambda^{\rm rf} = \lambda^{\rm rf}_{\rm 1}=1207.6$\,\AA, but this is not an issue since in our analysis we always have $\lambda^{\rm rf} < \lambda^{\rm rf}_{\rm 1}$. 

We show in figure \ref{fig:modfit} that in a direct fit to the redshift error contamination, our additional correction captures the trough feature very well. However, when performing a full-shape fit to our auto- and cross-correlations we find that $a_1$ takes much larger than expected values \footnote{Our expected value of $a_1$ is computed in a direct fit to the contamination, where the only free parameters are $\sigma_{\rm cont}$, $A_{\rm cont}$ and $a_1$.}. This consequently biased our measurements of $\phi_{\rm s}$, so it was decided to proceed for now without the additional correction.

\section*{Affiliations}
$^{1}$Institut de F\'{i}sica d’Altes Energies (IFAE), The Barcelona Institute of Science and Technology, Edifici Cn, Campus UAB, 08193, Bellaterra (Barcelona), Spain\\
$^{2}$Lawrence Berkeley National Laboratory, 1 Cyclotron Road, Berkeley, CA 94720, USA\\
$^{3}$Institut d'Astrophysique de Paris. 98 bis boulevard Arago. 75014 Paris, France\\
$^{4}$IRFU, CEA, Universit\'{e} Paris-Saclay, F-91191 Gif-sur-Yvette, France\\
$^{5}$Department of Physics, Boston University, 590 Commonwealth Avenue, Boston, MA 02215 USA\\
$^{6}$Dipartimento di Fisica ``Aldo Pontremoli'', Universit\`a degli Studi di Milano, Via Celoria 16, I-20133 Milano, Italy\\
$^{7}$INAF-Osservatorio Astronomico di Brera, Via Brera 28, 20122 Milano, Italy\\
$^{8}$Department of Physics \& Astronomy, University College London, Gower Street, London, WC1E 6BT, UK\\
$^{9}$Institute for Computational Cosmology, Department of Physics, Durham University, South Road, Durham DH1 3LE, UK\\
$^{10}$Instituto de F\'{\i}sica, Universidad Nacional Aut\'{o}noma de M\'{e}xico, Circuito de la Investigaci\'{o}n Cient\'{\i}fica, Ciudad Universitaria, Cd. de M\'{e}xico  C.~P.~04510,  M\'{e}xico\\
$^{11}$Department of Astronomy \& Astrophysics, University of Toronto, Toronto, ON M5S 3H4, Canada\\
$^{12}$Department of Physics \& Astronomy and Pittsburgh Particle Physics, Astrophysics, and Cosmology Center (PITT PACC), University of Pittsburgh, 3941 O'Hara Street, Pittsburgh, PA 15260, USA\\
$^{13}$Departamento de F\'isica, Universidad de los Andes, Cra. 1 No. 18A-10, Edificio Ip, CP 111711, Bogot\'a, Colombia\\
$^{14}$Observatorio Astron\'omico, Universidad de los Andes, Cra. 1 No. 18A-10, Edificio H, CP 111711 Bogot\'a, Colombia\\
$^{15}$Institut d'Estudis Espacials de Catalunya (IEEC), c/ Esteve Terradas 1, Edifici RDIT, Campus PMT-UPC, 08860 Castelldefels, Spain\\
$^{16}$Institute of Cosmology and Gravitation, University of Portsmouth, Dennis Sciama Building, Portsmouth, PO1 3FX, UK\\
$^{17}$Institute of Space Sciences, ICE-CSIC, Campus UAB, Carrer de Can Magrans s/n, 08913 Bellaterra, Barcelona, Spain\\
$^{18}$Fermi National Accelerator Laboratory, PO Box 500, Batavia, IL 60510, USA\\
$^{19}$Center for Cosmology and AstroParticle Physics, The Ohio State University, 191 West Woodruff Avenue, Columbus, OH 43210, USA\\
$^{20}$Department of Physics, The Ohio State University, 191 West Woodruff Avenue, Columbus, OH 43210, USA\\
$^{21}$The Ohio State University, Columbus, 43210 OH, USA\\
$^{22}$Department of Physics, The University of Texas at Dallas, 800 W. Campbell Rd., Richardson, TX 75080, USA\\
$^{23}$Department of Physics, Southern Methodist University, 3215 Daniel Avenue, Dallas, TX 75275, USA\\
$^{24}$Department of Physics and Astronomy, University of California, Irvine, 92697, USA\\
$^{25}$Sorbonne Universit\'{e}, CNRS/IN2P3, Laboratoire de Physique Nucl\'{e}aire et de Hautes Energies (LPNHE), FR-75005 Paris, France\\
$^{26}$Departament de F\'{i}sica, Serra H\'{u}nter, Universitat Aut\`{o}noma de Barcelona, 08193 Bellaterra (Barcelona), Spain\\
$^{27}$Department of Astronomy, The Ohio State University, 4055 McPherson Laboratory, 140 W 18th Avenue, Columbus, OH 43210, USA\\
$^{28}$Instituci\'{o} Catalana de Recerca i Estudis Avan\c{c}ats, Passeig de Llu\'{\i}s Companys, 23, 08010 Barcelona, Spain\\
$^{29}$Department of Physics and Astronomy, Siena College, 515 Loudon Road, Loudonville, NY 12211, USA\\
$^{30}$Departamento de F\'{\i}sica, DCI-Campus Le\'{o}n, Universidad de Guanajuato, Loma del Bosque 103, Le\'{o}n, Guanajuato C.~P.~37150, M\'{e}xico\\
$^{31}$Instituto Avanzado de Cosmolog\'{\i}a A.~C., San Marcos 11 - Atenas 202. Magdalena Contreras. Ciudad de M\'{e}xico C.~P.~10720, M\'{e}xico\\
$^{32}$Department of Physics and Astronomy, University of Waterloo, 200 University Ave W, Waterloo, ON N2L 3G1, Canada\\
$^{33}$Perimeter Institute for Theoretical Physics, 31 Caroline St. North, Waterloo, ON N2L 2Y5, Canada\\
$^{34}$Waterloo Centre for Astrophysics, University of Waterloo, 200 University Ave W, Waterloo, ON N2L 3G1, Canada\\
$^{35}$Instituto de Astrof\'{i}sica de Andaluc\'{i}a (CSIC), Glorieta de la Astronom\'{i}a, s/n, E-18008 Granada, Spain\\
$^{36}$Departament de F\'isica, EEBE, Universitat Polit\`ecnica de Catalunya, c/Eduard Maristany 10, 08930 Barcelona, Spain\\
$^{37}$Department of Physics and Astronomy, Sejong University, 209 Neungdong-ro, Gwangjin-gu, Seoul 05006, Republic of Korea\\
$^{38}$CIEMAT, Avenida Complutense 40, E-28040 Madrid, Spain\\
$^{39}$Department of Physics, University of Michigan, 450 Church Street, Ann Arbor, MI 48109, USA\\
$^{40}$University of Michigan, 500 S. State Street, Ann Arbor, MI 48109, USA\\
$^{41}$Department of Physics \& Astronomy, Ohio University, 139 University Terrace, Athens, OH 45701, USA\\
$^{42}$NSF NOIRLab, 950 N. Cherry Ave., Tucson, AZ 85719, USA\\
$^{43}$National Astronomical Observatories, Chinese Academy of Sciences, A20 Datun Road, Chaoyang District, Beijing, 100101, P.~R.~China\\

\bsp	
\label{lastpage}

\end{document}